\documentclass[lettersize,journal]{IEEEtran}
\usepackage{amsmath,amsfonts}
\usepackage[ruled,lined,linesnumbered]{algorithm2e}
\usepackage{array}
\usepackage{textcomp}
\usepackage{stfloats}
\usepackage{url}
\usepackage{verbatim}
\usepackage{cite}
\usepackage[colorlinks, citecolor=black]{hyperref}
\usepackage[capitalize]{cleveref}
\crefname{section}{Sec.}{Secs.}
\Crefname{section}{Section}{Sections}
\Crefname{table}{Table}{Tables}

\usepackage{ mathrsfs }
\usepackage{multirow,diagbox}
\usepackage{verbatim}
\usepackage{color}
\usepackage{colortbl}
\usepackage{makecell}
\usepackage{cellspace}
\usepackage{stfloats}
\usepackage{longtable}
\usepackage{booktabs}
\usepackage{array,multirow,multicol,graphicx}
\usepackage{color}
\usepackage{float}
\usepackage{makecell}
\usepackage[x11names,dvipsnames,table]{xcolor} 
\usepackage{colortbl}
\usepackage{multirow}
\usepackage[font=footnotesize]{caption}
\usepackage{subcaption}
\usepackage{lipsum}
\definecolor{mygray}{gray}{0.6}
\definecolor{myblue}{rgb}{0.8,0.85,1}
\usepackage{array}
\newcolumntype{L}[1]{>{\raggedright\let\newline\\\arraybackslash\hspace{0pt}}m{#1}}
\newcolumntype{C}[1]{>{\centering\let\newline\\\arraybackslash\hspace{0pt}}m{#1}}
\newcolumntype{R}[1]{>{\raggedleft\let\newline\\\arraybackslash\hspace{0pt}}m{#1}}

\usepackage{array, tabularx, boldline}
\usepackage{eurosym}
\usepackage{amstext} 
\usepackage[noend]{algpseudocode}
\DeclareRobustCommand{\officialeuro}{%
  \ifmmode\expandafter\text\fi
  {\fontencoding{U}\fontfamily{eurosym}\selectfont e}}
\usepackage{amsthm}
\theoremstyle{plain}
\newtheorem{theorem}{Theorem}
\newtheorem{assumption}{Assumption}
\usepackage[marginal]{footmisc}

\begin{document}

\title{Towards Efficient and Certified Recovery from Poisoning Attacks in Federated Learning}

\author{Yu Jiang$^{\dagger}$, Jiyuan Shen$^{\dagger}$, Ziyao Liu, Chee Wei Tan, and Kwok-Yan Lam

%
%
\thanks{Yu Jiang and Jiyuan Shen are with the School of Computer Science and Engineering (SCSE), Nanyang Technological University, Singapore. Ziyao Liu is with Digital Trust Centre (DTC), Nanyang Technological University, Singapore. Chee Wei Tan is with the School of Computer Science and Engineering (SCSE), Nanyang Technological University, Singapore.
Kwok-Yan Lam is with the School of Computer Science and Engineering (SCSE) and Digital Trust Centre (DTC), Nanyang Technological University, Singapore.  
E-mail: \{yu012, jiyuan001\}@e.ntu.edu.sg, liuziyao@ntu.edu.sg, 
cheewei.tan@ntu.edu.sg, kwokyan.lam@ntu.edu.sg.}
\thanks{Manuscript received November 15, 2023; revised January 16, 2024; accepted April 14, 2024.}}

\markboth{Journal of \LaTeX\ Class Files,~Vol.~14, No.~8, November~2023}%
{Shell \MakeLowercase{\textit{et al.}}: A Sample Article Using IEEEtran.cls for IEEE Journals}

\IEEEpubid{0000--0000/00\$00.00~\copyright~2021 IEEE}

\maketitle
\IEEEpubidadjcol
\footnote{$^{\dagger}$ Both authors contributed equally to this research.}
\begin{abstract}

Federated learning (FL) is vulnerable to poisoning attacks, where malicious clients manipulate their updates to affect the global model.
Although various methods exist for detecting those clients in FL, identifying malicious clients requires sufficient model updates, and hence by the time malicious clients are detected, FL models have been already poisoned. Thus, a method is needed to recover an accurate global model after malicious clients are identified. Current recovery methods rely on (i) all historical information from participating FL clients and (ii) the initial model unaffected by the malicious clients, both leading to a high demand for storage and computational resources.

In this paper, we show that highly effective recovery can still be achieved based on (i) selective historical information rather than all historical information and (ii) a historical model that has not been significantly affected by malicious clients rather than the initial model. In this scenario, we can accelerate the recovery speed and decrease memory consumption as well as maintaining comparable recovery performance. Following this concept, we introduce Crab, an efficient and certified recovery method, which relies on selective information storage and adaptive model rollback. Theoretically, we demonstrate that the difference between the global model recovered by Crab and the one recovered by train-from-scratch can be bounded under certain assumptions. 
Our empirical evaluation, conducted across three datasets over multiple machine learning models and a variety of untargeted and targeted poisoning attacks, demonstrates that Crab is not only accurate and efficient but also consistently outperforms previous approaches in terms of recovery speed and memory consumption.

\end{abstract}


\section{Introduction}
In recent years, federated learning (FL) \cite{kairouz2021advances, zhang2021survey, liu2022privacy, li2020federated, hang2023privacy} has emerged as a promising paradigm for collaborative and privacy-enhancing machine learning. In a typical FL system, data owners conduct machine learning (ML) individually on their local data, computing gradients that are subsequently aggregated to construct a global model. However, recent studies indicate that malicious clients can employ various poisoning attacks on the global model by compromising their local models through either (i) untargeted attacks aimed at slowing down the learning process or decreasing the overall performance of the global model \cite{shejwalkar2022back}, or (ii) targeted attacks (also known as backdoor attacks) that introduce a backdoor into a model, causing it to exhibit malicious behaviors when the inputs contain a predefined trigger \cite{jebreel2023fl,bagdasaryan2020backdoor}. Many poisoning attacks can swiftly impact the global model's performance or inject backdoors within a few FL rounds, persisting for a duration exceeding numerous rounds \cite{zhang2023fltracer}, thereby raising significant security concerns\cite{blanchard2017machine, lyu2020threats}. Hence, various methods \cite{zhang2022fldetector, li2020learning, shen2016auror} have been proposed to detect malicious clients either during or after the training process. However, these detection methods often require sufficient model updates to make confident decisions, and hence by the time malicious clients are identified, they may have already poisoned the global model. Consequently, there is a need for the server to recover an accurate global model from the poisoned one after the detection of malicious clients.

\IEEEpubidadjcol

\begin{figure}[t]
    \centering
    \includegraphics[width=1\linewidth]{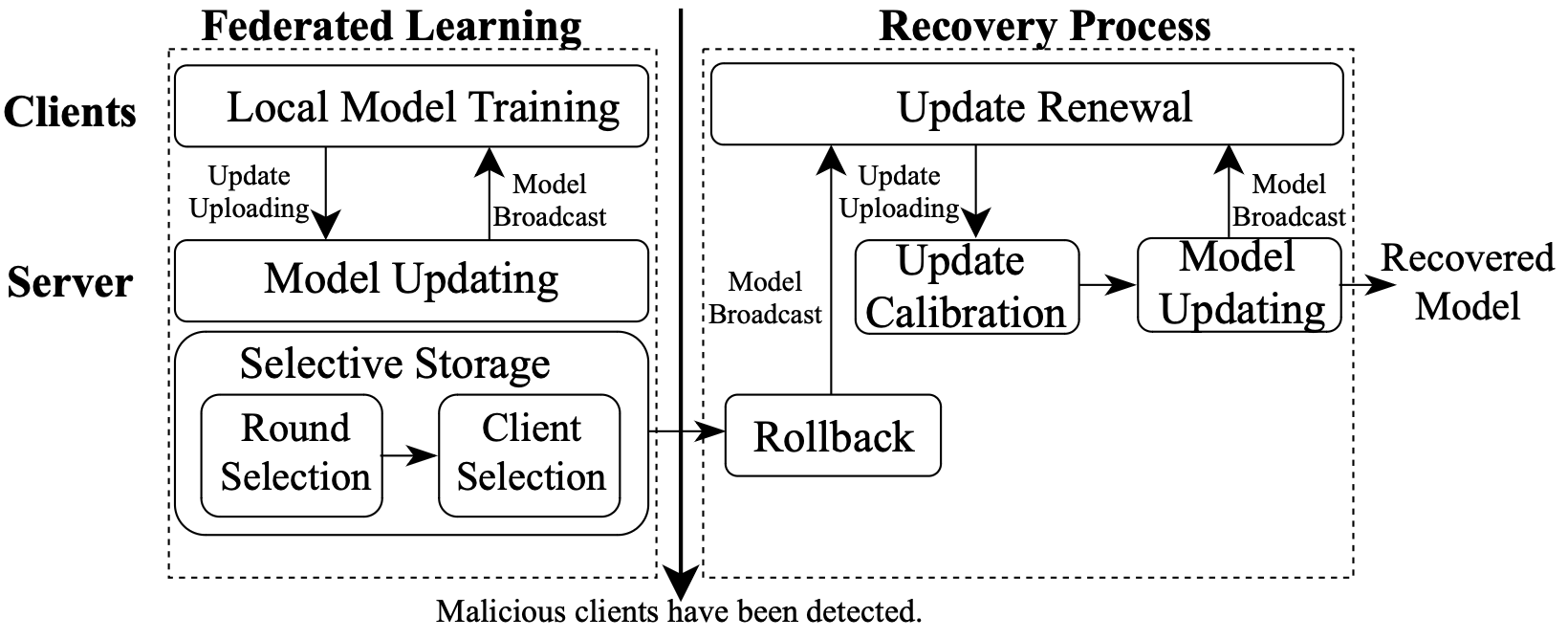}
    \caption{An illustration of Crab scheme.}
    \label{fig:workflow}
\end{figure}

To recover the FL model from poisoning attacks after detecting malicious clients, the server can simply remove the identified malicious clients and initiate the training of a new global model from scratch conducted by the benign clients. This straightforward recovery method, despite effective, imposes significant computation and communication costs on FL clients. This drawback may limit its applicability, especially in large-scale FL systems with resource-constrained clients. Current approaches achieve efficient recovery by mitigating the impact of malicious clients \cite{cao2023fedrecover,liu2021federaser,liu2023survey}, utilizing (i) all historical information from all participating FL clients and (ii) the initial model unaffected by the malicious clients. More specifically, the server stores all historical information during FL training, including the global model and clients' gradients. Upon identifying malicious clients, the server initiates a rollback to the initial global model. Based on this initial global model, one can calibrate the clients' historical gradients in the initial FL round and obtain a new global model. Calibrations are then iteratively conducted for each FL round. This procedure is repeated until all historical gradients are properly calibrated, indicating the assumed model recovery.

However, these recovery schemes that involve storing all historical information and relying solely on an unaffected model for global model recovery incur significant storage and computational costs. Meanwhile, we have observed that (i) historical information varies in importance to the global model, and (ii) malicious clients have not significantly affected the global model at the beginning of the FL training. Thus, selecting important historical information offers sufficient information for recovery. Furthermore, employing a historical global model that has not been too much affected by malicious clients can be adequate to recover a reasonably accurate global model from the poisoned one. The above indicates that highly effective recovery can still be achieved based on (i) selective storage of important historical global models and clients' gradients rather than all historical information, and (ii) adaptive rollback to a global model that has not been significantly affected by the malicious clients rather than the initial model. In this case, the recovery speed can be accelerated, and memory consumption can be reduced. Following this concept, we introduce Crab (an acronym for ``Certified Recovery from poisoning Attack" and a metaphor illustrating that our method shares regenerative capabilities with a crab after encountering attacks), an efficient and certified recovery method, which relies on selective information storage and adaptive model rollback.

Specifically, Crab implements selective information storage at the round level and the client level. In other words, the server selects which FL rounds exhibit a rapid change in the global model and which clients significantly impact the change in these selected rounds. Based on these selections, the server stores the corresponding historical information. After identifying the malicious clients, the server is tasked with determining an optimal rollback model concerning the malicious clients. In Crab, the server calculates the sensitivity of all stored historical models to the malicious clients and designates the most recent one of which the sensitivity does not surpass a pre-defined threshold, as the rollback model. Based on the selected historical information and the designated rollback model, the benign clients follow the standard workflow for the recovery process adapted for Crab, as demonstrated in \cref{fig:workflow}.

Theoretically, we demonstrate that the difference between the global model recovered by Crab and the one recovered by train-from-scratch has an upper bounded under certain assumptions. This indicates the effectiveness of recovery and the certified client removal provided by Crab. 
Empirically, our experiments over various datasets and different poisoning attacks settings can effectively demonstrate the superiority of proposed Crab in terms of both accuracy and efficiency. Besides, it consistently outperforms previous approaches in terms of recovery speed and memory consumption.


\textbf{Related works.} The idea of recovering a global model from a poisoned one in a federated learning setting is related to the concept of federated unlearning (FU) \cite{halimi2022federated,liu2021federaser,liu2022right}. Specifically, model recovery in federated learning can be conceptualized as the process of unlearning one or more detected malicious clients, involving making the global model forget the model updates contributed by these identified malicious clients. The works most closely related to Crab are FedEraser \cite{liu2021federaser} and FedRecover \cite{cao2023fedrecover}, both of which achieve unlearning or recovery based on historical information and the initial global model. A difference between these two works is that in FedEraser \cite{liu2021federaser}, the server stores historical information at intervals of every fixed number of rounds, whereas in FedRecover \cite{cao2023fedrecover}, the server stores historical information for all FL rounds. 
However, both FedEraser and FedRecover require high memory consumption and computational cost.
In contrast, Crab stores only important historical information from selected FL rounds and rolls back the model to a recent historical one, as opposed to the initial model. Crab has the capability to remarkably reduce storage space and computational cost, with only a marginal trade-off in performance.

\textbf{Our contributions.} The main contributions of this work are outlined below.
\begin{enumerate}
\item We introduce Crab, an efficient and certified method designed for recovering an accurate global model from poisoning attacks in federated learning.  
\item We propose two strategies, selective information storage and adaptive model rollback, tailored for model recovery to reduce memory consumption and expedite the recovery speed.
\item 
We offer a certified recovery by demonstrating the existence of a theoretical upper bound on the difference between the global model recovered by Crab and the one recovered by train-from-scratch under certain assumptions.
\item We conduct extensive experiments to evaluate the performance of Crab across three datasets over multiple machine learning models, and a variety of untargeted and targeted poisoning attacks, demonstrating its accuracy, cost-effectiveness, and efficiency.
\end{enumerate}

\textbf{Organisation of the paper.} Section \ref{sec:preliminaries} presents the preliminaries and introduces the notations used throughout the paper. In Section \ref{problem_definition}, we outline the threat model and discuss the design goals. Section \ref{sec:proposed_protocol} provides a detailed description of our proposed methods along with a theoretical analysis. The experimental evaluation is presented in Section \ref{sec:experimental_evaluation}. Finally, Section \ref{sec:conclusions} concludes the paper.

\section{Preliminaries and Notations}
\label{sec:preliminaries}

\subsection{Federated Learning}
In the standard architecture of federated learning, the participants can be divided into two main categories: (i) a set of $C$ clients, and
(ii) a central server denoted as $S$. 
The server $S$ trains a global model with training dataset distributed across the $C$ clients. Each client $c \in [C]$ possesses a local training dataset $D_c$, which construct the full training dataset $D=\{D_1, D_2, \cdots, D_C\}$.
The $C$ clients achieve the goal by minimizing a loss function, i.e., $\min_M f(D_c;M)$, where $M$ denotes the global model and $f$ is the empirical loss function, i.e., cross-entropy loss. For simplicity, we let $f_c(M)$ denote the loss function of client $c$ with dataset $D_c$,
and $F(M)$ denote the overall loss based on the full training dataset $D$.
Specifically, the server organizes the model training process, by repeating the following steps until the training is stopped \cite{kairouz2021advances}. At the $t$-th training round, the server broadcast the global model ${M_{t}}$ to all the clients, then the workflow is performed as follows:
\begin{enumerate}
    \item Local Model Training: Each federated learning client $c$ independently trains its local model to compute the update $\boldsymbol g_t^c$ applying the stochastic gradient descent based on its private local dataset ${D}_c$ after receiving global model ${M_{t}}$. 
    \item Model Uploading: Each client $c$ uploads its locally update $\boldsymbol g_t^c$ to the central server $S$. 
    \item Model Aggregation: The central server $S$ aggregates the received updates $\{\boldsymbol g_t^1,\boldsymbol g_t^2, \cdots,\boldsymbol g_t^C\}$ from all clients according to an aggregation rule $\mathcal{A}$ \cite{mcmahan2017communication, li2020federated2, zhang2023symmetric, wang2020tackling, karimireddy2020scaffold}, such that aggregated update $\boldsymbol G_t=\mathcal{A}(\boldsymbol g_t^1,\boldsymbol g_t^2, \cdots,\boldsymbol g_t^C)$. 
    \item Model Updating: The central server $S$ use the aggregated information to update the global model ${M_{t}}$ and distributes the updated global model ${M_{t+1}}$ to all federated learning clients. 
\end{enumerate}
The above iterative training process will be performed until some convergence criterion has been met or the training objective has been satisfied. Finally, the server will obtain the final model.



\subsection{Poisoning Attacks}
Federated learning is vulnerable to poisoning attacks, where malicious clients compromise their local models using either untargeted attacks or targeted attacks which are known as backdoor attacks.

\textbf{Trim attack.} 
Initially introduced in \cite{fang2020local}, the trim attack is an untargeted poisoning attack designed to disrupt the correct optimization trajectory of the global model in federated learning. It achieves this by carefully crafting malicious model updates. These updates may involve adding random Gaussian noise, replacing certain parameter values, or omitting specific values. While the trim attack was primarily developed for the Trimmed-mean aggregation rule, it has also proven effective with other aggregation methods, such as FedAvg \cite{mcmahan2017communication}.

\textbf{Backdoor attack.} 
Backdoor attacks in machine learning involve embedding a distinct pattern or ``trigger" into training data portions, as described in \cite{li2022backdoor, shejwalkar2022back}. This trigger could be a conspicuous element like a small patch \cite{gu2017badnets,liu2018trojaning}, or subtle alterations in benign samples that are less perceptible \cite{li2020invisible,bagdasaryan2021blind,saha2020hidden}. While the model behaves normally with standard inputs, it responds maliciously to inputs with this hidden trigger. The danger of backdoor attacks stems from their dormant nature, becoming active only under the attacker's control.

\subsection{Notations}
We summarize the parameters and notations used throughout the paper in Table \ref{tab:notations}.

\section{Problem Definition}\label{problem_definition}

\subsection{Threat Model}

We adopt the threat model presented in FedRecover \cite{cao2023fedrecover}, which aligns with the models used in many prior works \cite{bagdasaryan2020backdoor,bhagoji2019analyzing,fang2020local,xie2019dba}. Assuming the presence of an honest server in an FL system, the adversary can consist of a set of colluding active malicious clients, which may manipulate their updates, i.e., gradients or local models to be uploaded to the server, to compromise the performance of the FL system. Specifically, the adversary, armed with knowledge of data and model updates on malicious clients, or having access to information on all participating clients, can execute either untargeted poisoning attacks \cite{fang2020local,shejwalkar2021manipulating} aimed at reducing the FL global model's performance, or targeted poisoning attacks \cite{bagdasaryan2020backdoor,baruch2019little,bhagoji2019analyzing} designed to insert a backdoor into the model, leading it to misclassify data samples with a backdoor trigger as a class selected by the adversary.

\begin{table}[t!]
\centering
\caption{Notations of parameters used throughout the paper}
\label{tab:notations}
\begin{tabular}{c|l}
\toprule
Notation & Description \\ \midrule
$c$ & Client index\\
$b$ & Time window index \\
$t_j$ & Stored global model index\\ 
$\alpha$ & Percentage of loss reduction \\
$\beta$ & Sensitivity ratio \\
$\eta$ & Learning rate\\
$\kappa$ & Total number of time windows\\
$C$ & Total number of FL clients \\
$C_u$ & Malicious client set\\
$\lambda$ & Round selection ratio in each time window\\
$\delta$ & Client selection ratio for each selected model\\
$E$ & Total number of local training rounds from client $c$ \\
$T$ & Total number of learning rounds \\
$T^\prime$ & Total number of stored global models\\
$B_b$ & Buffer size of $b$-th time window\\
$D_c$ & Dataset of client $c$\\
$F(\cdot)$ & Loss function \\
$p(\cdot)$ & Data probability distribution \\
\bottomrule
\end{tabular}
\end{table}
\subsection{Design Goals}

We aim to design an efficient and certified model recovery method in an FL setting. Specifically, we outline the following design objectives:

\textbf{Certified recovery.} We use the train-from-scratch approach as a baseline to evaluate our model recovery method's performance. Theoretically, under certain assumptions, the difference between the model recovered by Crab and the baseline should be limited. Empirically, in the context of poisoning attacks, the evaluation metrics of recovery on the recovered model should align closely with that of the baseline.

\textbf{Cost-effective recovery.} The process of recovery via Crab should be cost-effective, with a specific focus on minimizing memory consumption for the storage of historical information. This emphasis arises from the challenge faced in large-scale FL systems, where storing extensive historical data can lead to significant memory consumption. Our goal is to design a cost-effective recovery method that necessitates the server to maintain only a compact subset of historical information.

\textbf{Fast recovery.} The number of rounds needed for recovery using Crab should be fewer compared to the train-from-scratch baseline. We introduce the average round-saving percentage in Section \ref{sec:experimental_evaluation} to measure the speed of recovery. A recovery method is considered faster if it has a higher average round-saving percentage.



\section{Design of Crab}
\label{sec:proposed_protocol}

\subsection{Overview}
As described earlier, in the state-of-the-art scheme FedRecover \cite{cao2023fedrecover}, the server stores all historical global models and clients' gradients during the FL training process. After the malicious clients are detected, the server rolls back the global model to the initial one, from which recovery is conducted based on stored historical information. Compared to FedRecover \cite{cao2023fedrecover}, our proposed scheme, Crab, ensures a more efficient recovery through two key strategies: (i) selective storage of important historical information rather than all historical information, and (ii) adaptive rollback to a global model that has not been significantly affected by the malicious clients rather than the initial model. The overall Crab scheme is shown in Fig. \ref{fig:workflow}. In this section, we will present the detailed designs for these two strategies within the full description of the Crab scheme. Additionally, we will provide a theoretical analysis of the difference between the global model recovered by Crab and the one obtained through train-from-scratch.



\subsection{Design Details}
\subsubsection{Selective Information Storage} 
To minimize memory consumption, we implement two strategies for the selective storage of historical information at (i) the round level and (ii) the client level.

\textbf{Round selection.} We observe that throughout the FL training process, the global model converges at varying rates in different rounds. A larger rate indicates a more significant change in the global model, emphasizing the greater importance of the updates, i.e., the gradients, are more important for the global model. Therefore, selectively storing these important gradients allows for a substantial reduction in memory consumption while preserving the performance of the unlearning algorithm. 
We utilize Kullback-Leibler (KL) divergence to measure the difference between a global model and the global model in the subsequent FL round.
KL divergence is a measure of how one probability distribution diverges from a second probability distribution and used to quantify the difference between two probability distributions \cite{cover1999elements}.
Comparing the KL divergence between two models can provide insights into how much the models differ from each other. Besides, KL divergence has been a widely employed metric for evaluating the distinction between two ML models \cite{fang2022robust, zhang2022personalized}.  A higher KL divergence indicates a more significant change in the global model, emphasizing the greater importance of the gradients in that FL round, which are then selected for storage with a higher probability.

To implement round selection based on KL divergence during FL training, we instruct the server to buffer, i.e., store in a temporary memory, historical gradients from all FL rounds within a specified time window. The server then assesses the KL divergence of these stored historical gradients with the global model, distinguishing between important gradients to store and non-important ones to discard. After that, the selection process moves on to the next specified time window. The calculation of the time window is based on achieving a fixed percentage reduction in the loss of the global model.

In particular, We designate $T$ as the total number of FL training rounds conducted prior to the identification of malicious clients. The loss of the global model $M_t$ at the $t$-th FL training round is denoted by $F(M_t)$. We then define the loss reduction rate as $\gamma = \frac{F(M_0) - F(M_{T-1})}{F(M_0)}$, where $F(M_0)$ and $F(M_{T-1})$ represent the losses of the global models at the initial and final rounds, respectively. Furthermore, the global models are stored within a set of time windows in a buffer. If there are $B$ such time windows across the $T$ FL rounds, the loss of the last global model in each time window is reduced by a rate of $\alpha$ compared to the loss of the first global model in the same time window. Therefore, we calculate $B$ as $B= \left\lfloor \frac{\log(1-\gamma)}{\log(1-\alpha)} \right\rfloor$, where $\left\lfloor \cdot \right\rfloor$ denotes the floor function.


Specifically, let $B_b$ represent the size of the $b$-th time window, with $b$ belonging to the set $[B]$. In this $b$-th time window, the server temporarily stores consecutive global models along with their gradients in the buffer, denoted as $\{M_{b,i}\}^{B_b}_{i=1}$. Consequently, the loss reduction percentage between the first model $M_{b,1}$ and the last model $M_{b,B_b}$ in each time window is $\alpha$, satisfying the equation $F(M_{b,B_b}) = (1-\alpha)F(M_{b,1})$. Furthermore, let $p(M_{b,i})$ denote the data probability distribution of the full training dataset $D$ in the global model $M_{b,i}$. For each global model $M_{b,i}$ stored in the buffer, the server calculates the divergence from its subsequent model, $M_{b,i+1}$, using KL divergence, thereby quantifying the variation between successive models:
\begin{equation}
\begin{aligned}
       \mathcal{D}(M_{b,i},M_{b,i+1})
      &= KL[p(M_{b,i+1})|| p(M_{b,i})] \\
      &= \sum p(M_{b,i}) \cdot \log \frac{p(M_{b,i})}{p(M_{b,i+1})}.
\end{aligned}
\end{equation}
Subsequently, the server selects $\lambda$ percentage of total rounds that exhibit the highest values of $\mathcal{D}(M_{b,i},M_{b,i+1})$, which is the divergence between consecutive global models. It stores the gradients from these selected FL rounds and discards the others from the buffer. The stored global models and gradients are then represented as $\{\bar{M}_{b,i}\}^{P_b}_{i=1}$, where $P_b=\lambda B_b$. Throughout the FL training process, the number of the total stored global models is $T^{\prime}=\lambda T$. Thus, the stored global models can be presented as $\{\bar M_{t_j}\}^{T^\prime}= \{\bar M_{t_1}, \bar M_{t_2},\cdots,\bar M_{t_{T^\prime}} \}$. Correspondingly, the update is $ \{G_{t_j}\}^{T^\prime}=\{G_{t_1},G_{t_2},\cdots,G_{t_{T^\prime}}\}$ from all $C$ clients, where $G_{t_j}= \{\boldsymbol g_{t_j}^c\}_{c=1}^C$. A detailed description of this process is provided in Algorithm \ref{alg:roundSelection}.

\begin{algorithm*}
\SetAlgoNoEnd
\caption{Selective storage (SelectStore)}
\label{alg:roundSelection}
\KwIn{Buffer stored global model $M_{b,i}$ with buffer size $B_b$, each client $c$'s update $\boldsymbol g_{b,i}^c$, the aggregated update from $C$ clients $\boldsymbol G_{b,i}$, round selection ratio $\lambda$, client selection ratio $\delta$}
\KwOut{Selected global model $\{\bar{M}_{b,i}\}^{P_b}_{i=1}$ with aggregated updates $\{\boldsymbol{\bar G}_{b,i}\}^{{P_b}}_{i=1}$ from the high-contribution clients' updates $\{\boldsymbol{\bar g}_{b,i}^c\}_{c=1}^X$}
initial an empty set $\mathcal{D}$, $\mathcal{T}$;\\
\For{$i \leftarrow 1$ \KwTo $B_b$}{
calculate the model difference $\mathcal{D}(M_{b,i},M_{b,i+1})$; 
\Comment{\textbf{Round selection}}
\\
$\mathcal{D} \gets \mathcal{D} \cup {\mathcal{D}(M_{b,i},M_{b,i+1})}  $;
}
$\{\bar{M}_{b,i}\}^{P_b}_{i=1} \leftarrow $ SelectMax ($\mathcal{D}$, $\lambda B_b$);
\Comment{Select $P_b=\lambda B_b$ global models with the highest value of model difference}\\
\For{$i \leftarrow 1$ \KwTo $P_b$}{
\For{$c \leftarrow 1$ \KwTo $C$}{
calculate the contribution score $\mathcal{T} (\boldsymbol g_{b,i}^c, \boldsymbol G_{b,i})$;
\Comment{\textbf{Client selection}}\\
$\mathcal{T} \gets \mathcal{T} \cup \mathcal{T} (\boldsymbol g_{b,i}^c, \boldsymbol G_{b,i})$;
}
$\{\boldsymbol{\bar g}_{b,i}^c\}_{c=1}^X \leftarrow $ SelectMax ($\mathcal{T}$, $\delta C$)  
\Comment{Select $X=\delta C$ clients with the highest contribution scores}\\
$\boldsymbol{\bar G}_{b,i}= \mathcal{A}(\boldsymbol{\bar g}_{b,i}^1, \boldsymbol{\bar g}_{b,i}^2, \cdots, \boldsymbol{\bar g}_{b,i}^X)$ 
\Comment{Aggregated update from high-contribution clients}
\\
}
\KwRet{$\{\bar{M}_{b,i}\}^{P_b}_{i=1}$, $\{\boldsymbol{\bar G}_{b,i}\}^{{P_b}}_{i=1}$, $\{\boldsymbol{\bar g}_{b,i}^c\}_{c=1}^X$}
\end{algorithm*}

\textbf{Client selection.} Similar to the underlying concept of round selection, we have also observed that for a specific FL training round, clients contribute at varying levels. A larger contribution indicates a more significant impact on the global model updating, emphasizing the greater importance of the client for the global model. Therefore, selectively storing gradients from these important clients allows for a further reduction in memory consumption while preserving the performance of the unlearning algorithm. We employ cosine similarity to assess the direction similarity between a client's update, i.e., gradients, and the global update, i.e., the aggregated gradients from all participating clients, in the same FL round, a widely employed metric for evaluating the similarity between two vectors in an FL setting \cite{cao2021fltrust,liu2021privacy, zhang2023fltracer, ma2022shieldfl}. A higher Cosine similarity indicates a more substantial contribution of the client to the global model, indicating a greater importance of the clients in that FL round, which are then selected for storage with a higher probability.

After finalizing the storage of global models and updates through round selection, we proceed with client selection.
For the selected round $\{\bar M_{t_j}\}^{T^\prime}$, the update from $C$ clients is $\{G_{t_j}\}^{T^\prime}$, where $G_{t_j}= \{\boldsymbol g_{t_j}^c\}_{c=1}^C$, and the aggregated update is $\boldsymbol G_{t_j}= \mathcal{A}(\boldsymbol g_{t_j}^1, \boldsymbol g_{t_j}^2, \cdots, \boldsymbol g_{t_j}^C)$.
To evaluate the contribution of each client $c_i$, we use the cosine similarity to compute its contribution score as:
\begin{equation}
    \begin{aligned}
        \mathcal{T} (\boldsymbol g_{t_j}^c, \boldsymbol G_{t_j})
        &=  \frac{\left\langle \boldsymbol g_{t_j}^c, \boldsymbol G_{t_j} \right\rangle}{\Vert \boldsymbol g_{t_j}^c \Vert \cdot \Vert \boldsymbol G_{t_j} \Vert},
    \end{aligned}
\end{equation}
where $\left\langle  , \right\rangle$ is the scalar product operator.
Afterward, the server selects $\delta$ percent of total clients with the highest value of $ \mathcal{T} (\boldsymbol g_{t_j}^c, \boldsymbol G_{t_j})$ and stores the corresponding gradients. Finally, we can obtain the updates from the $C_{t_j}$ selected high-contribution clients represented as $\bar G_{t_j}= \{\boldsymbol{\bar g}_{t_j}^c\}_{c=1}^X$, and the aggregated update represented as $\boldsymbol{\bar G}_{t_j}= \mathcal{A}(\boldsymbol{\bar g}_{t_j}^1, \boldsymbol{\bar g}_{t_j}^2, \cdots, \boldsymbol{\bar g}_{t_j}^X)$, where $X= \delta C$.

\subsubsection{Adaptive Model rollback} 

In federated learning, to recover a global model compromised by poisoning, previous studies \cite{liu2021federaser, cao2023fedrecover} revert the model to a prior state unaffected by malicious clients, typically the initial model. Recovery then proceeds from this point. However, when the FL training involves many iterations, this recovery approach also requires numerous iterations, leading to significant computation and communication expenses. Fortunately, we find that rolling back to a historical model that has not been affected too much by malicious clients and conducting recovery from there is still effective. In this scenario, the recovery performance remains comparable, but the number of iterations required for the recovery process is substantially reduced, thereby significantly enhancing efficiency.

To assess a client's impact on the global model, we apply sensitivity analysis, a method often used in privacy and security for measuring how input variations affect outputs \cite{dwork2014algorithmic, fraboni2022sequential, guo2019certified}. In federated learning, higher sensitivity signifies a more significant influence of user parameters on the global model. To assess the sensitivity of each client, we first calculate the influence of client set $\boldsymbol{\bar g}_{t_j}^{c}$ to the global model as:
\begin{equation}
    \mathcal{I} (C_{t_j}) = \sum_{c=1}^X \frac{|D_c|}{| D_{{t_{j}}}|}  \cdot [\boldsymbol{\bar g}_{t_{j}}^{c} - \boldsymbol{\bar G}_{t_{j-1}}],
\end{equation}
where $\boldsymbol{\bar g}_{t_j}^{c}$ is the local update sent by client $c \in C_{t_j}$ after performing local training based on the global model $\bar M_{t_{j-1}}$, $D_{{t_{j}}}$ represents the dataset from client set $C_{t_j}$, and $ |\cdot|$ represents the size of a dataset.


After the $C_u$ malicious clients have been detected, we can calculate the sensitivity of one global model to the malicious clients:
\begin{equation}
    \begin{aligned}
        \mathcal{S}(C_u,j) = \sum_j \Vert \mathcal{I} (C_{t_j}) -\mathcal{I} (C_{t_j}^{-}) \Vert_2,
    \end{aligned}
\end{equation}
where $C_{t_j}^{-} = C_{t_j} \backslash C_u$ is benign client set obtained by removing the $C_u \in [1,C)$ malicious clients.

To identify the optimal rollback model, we set a threshold that helps in determining the most suitable model for rollback as follows:
\begin{equation}
    \Phi(j)= \beta \sum_j \mathcal{I} (C_{t_j}^{-}),
\end{equation}
where $\beta$ represents the sensitivity ratio, defining the sensitivity threshold as the $\beta$ fraction of the accumulated influence from benign clients. Therefore, according to our intuition of the adaptive sensitivity threshold, we can obtain the optimal roll-back model by solving the following problem:
\begin{equation}
    j^\star = \arg \max_{j} \left( \mathcal{S}(C_u,j) \leq \Phi(j) \right).
\end{equation}
Hence, we can obtain the optimal round $t_j=t_{j^\star}$ with corresponding roll-back model $\bar M_{t_{j^\star}}$.

Recall that during client selection, only historical information from high-contributing clients is stored. Consequently, there might be instances where the data from the detected $C_u$ malicious client were not chosen for storage in some stored rounds, leading to their model sensitivity being $0$. This scenario suggests that during FL training, the contribution and impact of the $C_u$ malicious clients on the model are negligible. This observation supports the notion that rolling back to the initial model may not always be necessary.

\begin{algorithm*}
\SetAlgoNoEnd
\caption{Crab Scheme}
\label{alg:Crab}
\KwIn{Federated learning client $c \in [C]$ with dataset $D_c$, initial Model $M_0$, the percentage of loss reduction $\alpha$, round $T$ before detecting the malicious clients, total stored round $T^\prime$, buffered time window $b$, time window size $B_b$, round selection ratio $\lambda$ for each time window, client selection ratio $\delta$ for each global model,
malicious clients $C_u$, pre-defined sensitivity threshold $\Phi(j)$}
\KwOut{Recovered global model $\tilde M$}
$\text{prev\_loss} \gets F(M_0)$; $b \gets 1$; $i \gets 0$;\\
\SetKwFunction{FL}{\textbf{Stage I. Federated Learning with Selective Storage}}
\SetKwProg{Fn}{}{:}{\KwRet}
\Fn{\FL}{
\For{$t \leftarrow 0$ \KwTo $T-1$}{
\For{$c \leftarrow 1$ \KwTo $C$ in parallel}{
client $c$ computes the update $\boldsymbol{g}_t^c$ by training locally on private dataset $D_c$ based on the global model $M_t$;
}
initial an empty set $M_b$; \\ $i \gets i+1$;\\
server $S$ aggregates the update $\boldsymbol G_t \gets \mathcal{A}(\boldsymbol g_t^1,\boldsymbol g_t^2, \cdots,\boldsymbol g_t^C)$;\\
server $S$ updates the global model $M_{t+1} \gets M_t-\eta \boldsymbol G_t$;\\
$M_{b,i} \gets M_{t+1}$;\\
server $S$ stores temporarily in the buffer for global model $M_b \gets M_b \cup \{M_{b,i}\}$ with corresponding clients' updates and the aggregated update;\\
\If{$F(M_{t+1}) \leq (1-\alpha) \cdot \text{prev\_loss}$}{
server $S$ computes the size $B_b$ for the $b$-th time window;\\
server $S$ update storage $\{\bar{M}_{b,i}\}^{P_b}_{i=1}, \{\boldsymbol{\bar G}_{b,i}\}^{P_b}_{i=1}, \{\boldsymbol{\bar g}_{b,i}^c\}_{c=1}^X \gets SelectStore(M_{b,i}, B_b, \boldsymbol g_{b,i}^c, \boldsymbol G_{b,i}, \lambda, \delta)$;\\
$b \gets b+1$;\\ $i \gets 0$;\\
$\text{prev\_loss} \gets F(M_{t+1})$;\\
}
}
At the end of round $T$, server $S$ stored $\{\bar M_{t_j}\}^{T^\prime}, \{\boldsymbol{\bar G}_{t_j}\}^{T^\prime}, \{\boldsymbol{\bar g}_{t_j}^c\}_{c=1}^C$;
}
\SetKwFunction{RB}{\textbf{Stage II. Adaptive Model Roll-back upon detecting the malicious clients $C_u$}}
\SetKwProg{Fn}{}{:}{\KwRet}
\Fn{\RB}{
server $S$ calculates the sensitivity $\mathcal{S}(C_u,j)$; \\
server $S$ finds the optimal roll-back model $\bar M_{t_{j^\star}}$ by solving the problem $\arg \max_{j} \left( \mathcal{S}(C_u,j) \leq  \Phi(j) \right)$;\\
}
\SetKwFunction{FU}{\textbf{Stage III. Recovery Process from Malicious Clients }}
\SetKwProg{Fn}{}{:}{\KwRet}
\Fn{\FU}{
$\tilde M_0=\bar M_{t_{j^\star}}$;\\
\For{$r \gets 0$ \KwTo $T^\prime-t_{j^\star}-1$}{
\For{$c \gets 1$ \KwTo $C \backslash C_u$ in parallel}{
client $c$ renews the update $\hat U_r^c$ by locally training on the private dataset $D_c$ based on previous recovered global model $\tilde M_r$;\\
}
server $S$ calibrates the update $\tilde U_r^c \gets \Vert U_{t_{j^\star + r }}^c \Vert \cdot \frac{\hat U_r^c}{\Vert \hat U_r^c \Vert}$;\\
server $S$ aggregates the update $\tilde U_r \gets \mathcal{A} (\tilde U_r^1,\tilde U_r^2,\tilde U_r^3,\cdots)$;\\
server $S$ updates the model $\tilde{M}_{r+1}  \gets \tilde{M}_{r}+ \tilde{U_r}$;\\
}
}
\KwRet{$\tilde M$}
\end{algorithm*}

\subsubsection{Model Recovery}

\begin{figure}[t]
    \centering
    \includegraphics[width=1\linewidth]{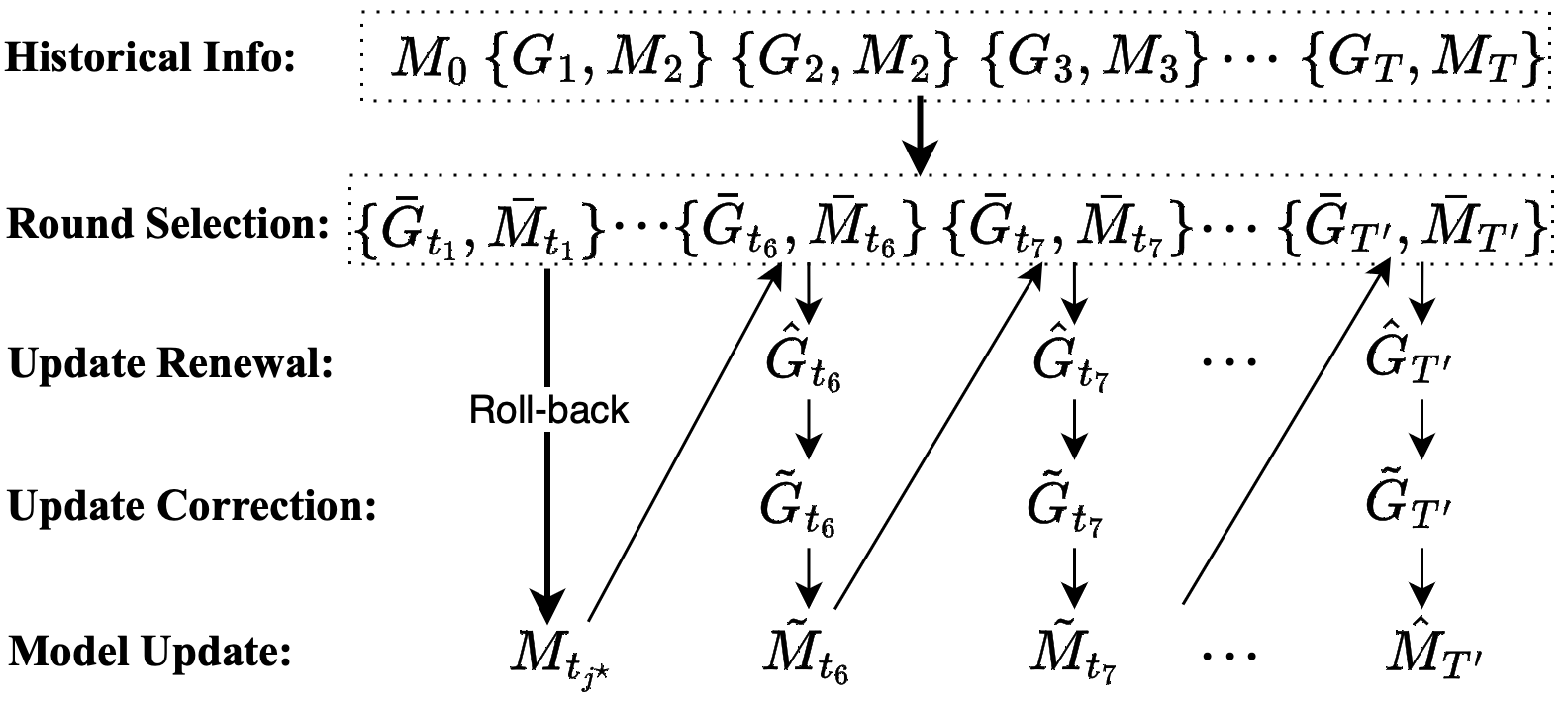}
    \caption{An illustration of the recovery process.}
    \label{fig:recovery_process}
\end{figure}

Once the optimal rollback model is identified, the benign clients and the server can proceed with the recovery task following the standard federated learning workflow based on the selective information storage, as depicted in \cref{fig:recovery_process}. Given the roll-back model $\bar M_{t_{j^\star}}$, the server is required to instruct the benign clients to do $T^\prime - j^\star$ rounds recovery. This involves four steps: (i) update renewal, (ii) update calibration, (iii) update aggregation, and (iv) model update. The first step is executed by the benign clients, while the server carries out the other three steps.

\textbf{Update renewal.} 
During the recovery process, at the $r$-th recovery round, each benign client $c \in [C \backslash C_u]$ locally trains $E$ rounds to compute the renewal update $\hat U_r^c$, based on the recovered global model $\tilde M_{r}$ from the previous round, which is then sent to the server:
\begin{equation}
\begin{aligned}
        \hat U_r^c &= M^c_E - M^c_0 \\
        &= (\tilde M_{r}- \eta \sum_{e=1}^E \hat{\boldsymbol g}_{r,e}^c)- \tilde M_{r}\\
        &= - \eta \sum_{e=1}^E \hat{\boldsymbol g}_{r,e}^c ,
\end{aligned}
\end{equation}
where $\hat{\boldsymbol g}_e^c$ is the update computed by the client $c$ in the $e$-th local training round, and $\tilde M_{r}$ is the initial model for the $r$-th round for the recovery process. Then each benign client $c$ sends the renewal update $\hat U_r^c = - \eta \hat{\boldsymbol g}_r^c$ to the server, where $\hat{\boldsymbol g}_r^c = \sum_{e=1}^E \hat{\boldsymbol g}_{r,e}^c$ represent the update computed by the client $c$ for the $r$-th recovery round.


\textbf{Update calibration.} 
Based on the renewal update $\hat U_r^c$ and the stored historical information, the server can calibrate the historical update. Let $U_{t_j}^c$ denote the historical update. We use the corresponding historical updates to calibrate the current renewal updates $\hat U_r^c$. Then, for $c \in [C^{-}_{t_{j^\star + r }}]$, we can use the norm of $U_{t_{j^\star + r }}^c$ to signify how much the parameters of the global model need to be changed, while the normalized $\hat U_r^c$ indicates in which direction the global model should be updated. Thus, the calibrated update can be obtained as:
\begin{equation}
\begin{aligned}
        \tilde U_r^c = \Vert U_{t_{j^\star + r }}^c \Vert \cdot \frac{\hat U_r^c}{\Vert \hat U_r^c \Vert} \\
        = - \eta \hat{\boldsymbol g}_r^c  \cdot \frac{\Vert \boldsymbol g_{t_{j^\star + r }}^c \Vert}{\Vert \hat{\boldsymbol g}_{r}^c \Vert}.
\end{aligned}
\end{equation}


\textbf{Update aggregation.} 
Once the server collects the calibrated updates, the server can aggregate them using an aggregation rule $\mathcal{A}$. FedAvg \cite{zhou2021communication} algorithm is employed in the Crab scheme for aggregation.

\begin{equation}
\begin{aligned}
    \tilde U_r&= \mathcal{A} (\tilde U_r^1,\tilde U_r^2,\tilde U_r^3,\cdots) \\
    &=\frac{1}{|D_{t_{j^\star + r }}^{-}|} \sum_{c \in [C_{t_{j^\star + r }}^{-}] } |D_c| \cdot U_r^c,
\end{aligned}
\end{equation}
where $D_{t_{j^\star + r }}^{-}$ represents the dataset from client set $C_{t_{j^\star + r }}^{-}$.


\textbf{Model update.} Based on the aggregated updates and the calibrated global model $\tilde{M}_{r}$ obtained from the previous recovery round, the calibrated global model for the subsequent recovery round can be obtained as:
\begin{equation}
\begin{aligned}
       \tilde{M}_{r+1}&=\tilde{M}_{r}+ \tilde{U_r}\\
       &=\tilde{M}_{r} - \eta  \sum_{c \in [C_{t_{j^\star + r }}^{-}] } \frac{ |D_c| }{|D_{t_{j^\star + r }}^{-}|}  \cdot \frac{\Vert \boldsymbol g_{t_{j^\star + r }}^c \Vert}{\Vert \hat{\boldsymbol g}_{r}^c \Vert} \cdot \hat{\boldsymbol g}_r^c  .
\end{aligned}
\end{equation}
The server and benign clients repeat the aforementioned four steps until all the historical gradients $U_{t_j}^c$ are calibrated. This process leads to the acquisition of the assumed recovered global model $\tilde{M}$. A detailed description of the Crab scheme is provided in Algorithm \ref{alg:Crab}.

\subsection{Theoretical Analysis}

In this section, we first outline the assumptions for our theoretical analysis. Then we demonstrate that the difference between the Crab model and the train-from-scratch method is bounded. A detailed deduction can be found in Appendix \ref{Proof of Theorem}.

\begin{assumption}[Lipschitz continuity]\label{ass_continuity}
    A function $F: \mathbb{R}^d \rightarrow \mathbb{R}$  is $L$-Lipschitz continuous, i.e.,$\forall M, M^\prime$, the following equation holds:
    \begin{equation}
        \Vert F(M) - F(M^\prime) \Vert \leq L \Vert M - M^\prime \Vert.
    \end{equation}
\end{assumption}
\begin{assumption}[L-smooth]\label{ass_smooth}
    A function $F: \mathbb{R}^d \rightarrow \mathbb{R}$  is $L$-smooth if it is differentiable and its gradient function $\nabla f:\mathbb{R}^d \rightarrow \mathbb{R}^d$ is Lipschitz continuous with constant $L$, i.e.$\forall M, M^\prime \in \mathbb{R}^d$, 
    \begin{equation}
        F(M^\prime) \leq F(M) + \nabla F(M)^\top (M^\prime - M) + \frac{L}{2} \Vert M^\prime - M \Vert^2.
    \end{equation}
\end{assumption}

\begin{assumption}[Bounded and unbiased Gradient]\label{ass_gradient}
    The stochastic gradient $f_c(M)$ has upper bound $G$ and is an unbiased estimator of federated loss function $f(M)$ \cite{beck2017first}:
    \begin{equation}
        \Vert f_c(M) \Vert \leq G.
    \end{equation}
\end{assumption}

\begin{theorem}[Model difference between Crab and train-from-scratch]\label{theo_1}
    The difference between the global model recovered by Crab in round $t$ and that recovered by train-from-scratch in round $\tau$ can be bounded as follows:
    \begin{equation}
\begin{aligned}
    &\Vert \tilde{M}_r -  M_{\tau} \Vert\\
    \leq & \sqrt{ \eta [ (F(\tilde M_0) - F(M^*))  ( r + \tau) + \frac{L}{2} \eta^2 G^2 (r\sum_{r=0}^{r-1} \check{\sigma}_r^2 +\tau) ]},
\end{aligned}
\end{equation}
where $\tilde{M}_r$ and $M_{\tau}$ are the global models recovered by Crab in round $r$ and train-from-scratch in round $\tau$ respectively, $\tau = \lceil r \cdot \frac{T}{(T^\prime-j^\star)} \rceil $, $\tilde M_0$ is the initial model used in both Crab and train-from-scratch method, $M^*$ is the optimal solution for the function $F(M)$, $\check{\sigma}_r = \sum_{c=1}^{X_{j^\star+r}} \sigma_r^c$, and $\sigma_r^c = \frac{ |D_c| }{|D_{t_{j^\star + r }}^{-}|}  \cdot \frac{\Vert \boldsymbol g_{t_{j^\star + r }}^c \Vert}{\Vert \hat{\boldsymbol g}_{r}^c \Vert} $.
\end{theorem}


Based on Theorem \ref{theo_1}, an upper bound is identified for the difference between the global model recovered via Crab and that derived from training from scratch, affirming the viability of our approach. Notably, there are trade-offs involved in selective storage and model rollback: choosing a more recent model for rollback often enlarges the difference bound, while storing fewer historical models may compromise recovery accuracy, and vice versa.

\begin{figure*}
    \centering
    \begin{subfigure}{0.8\linewidth}
        \centering
        \includegraphics[width=\linewidth]{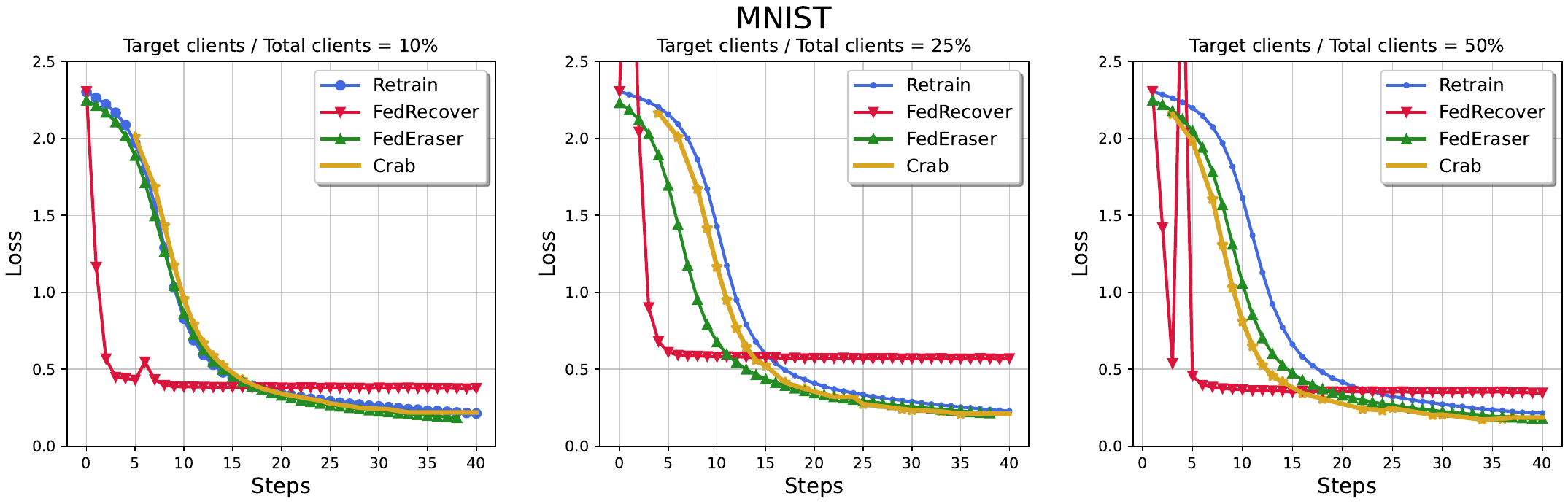}
        \label{fig:mnist_loss_vary}
    \end{subfigure}
    
    \begin{subfigure}{0.8\linewidth}
        \centering
        \includegraphics[width=\linewidth]{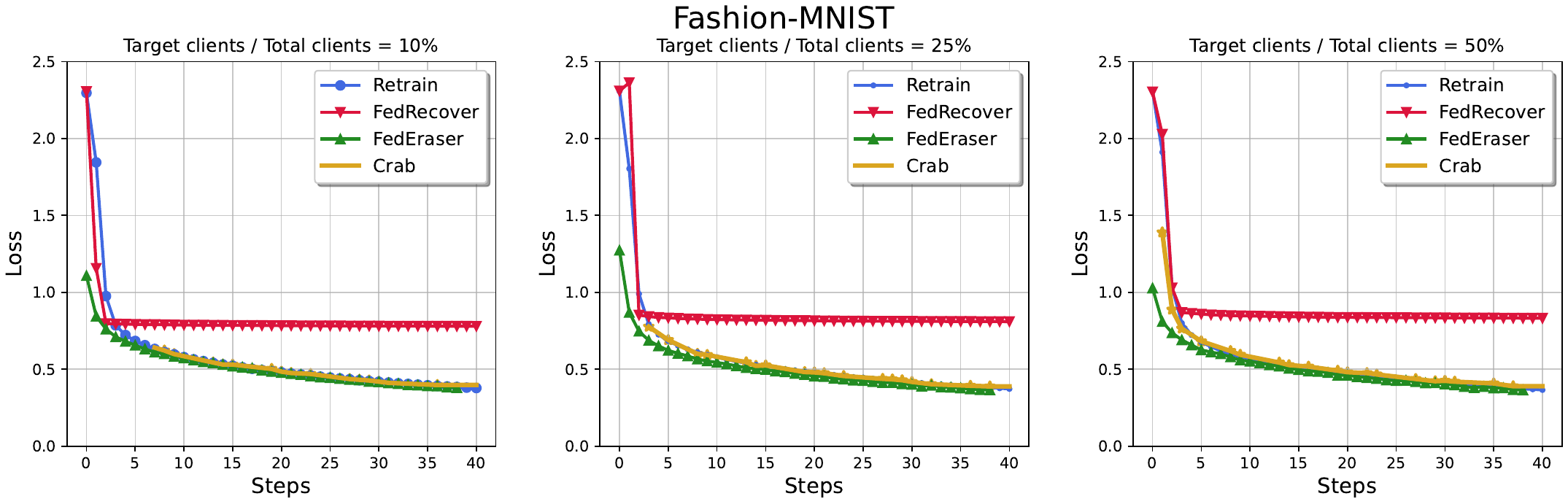}
        \label{fig:fmnist_loss_vary.}
    \end{subfigure}
    
    \begin{subfigure}{0.8\linewidth}
        \centering
        \includegraphics[width=\linewidth]{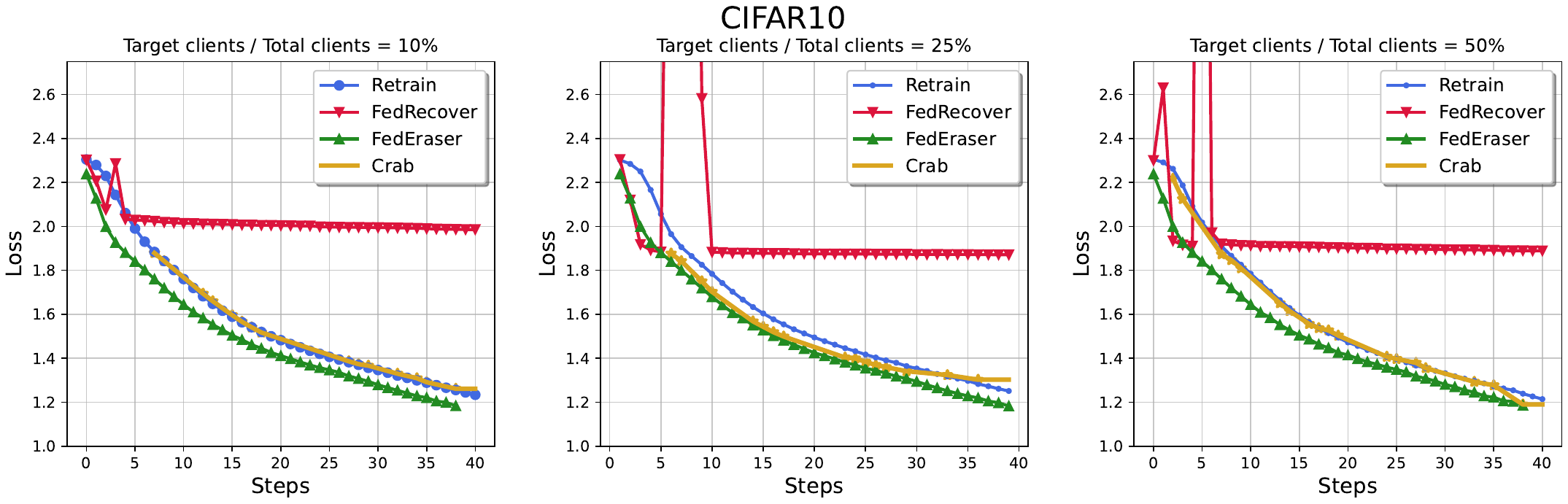}
        \label{fig:cifar10_loss_vary.}
    \end{subfigure}

    \caption{The loss descent curve of the recovery process on MNIST, Fashion-MNIST and CIFAR-10 when the percentage of target clients is 10\%, 25\%, 50\% respectively.}
    \label{fig:loss_descent}
\end{figure*}

\section{Evaluation} \label{sec:experimental_evaluation}
\subsection{Evaluation Setup}




\subsubsection{Implementation Details} We now proceed to outline the evaluation settings in terms of FL training, poisoning attack, and recovery parameters used in Crab.


\textbf{Federated learning settings.} We employ FedAvg \cite{mcmahan2017communication} as our aggregation method in the simulated FL environment comprising 20 clients. Each client performs 5 local training rounds, and in total, we conduct 40 global epochs. The clients utilize stochastic gradient descent (SGD) without momentum as their optimizer. A uniform learning rate of 0.005 and a batch size of 64 are set for these simulations.

\textbf{Attack settings.} We executed three levels of attack intensity, randomly designating 10\%, 25\%, and 50\% of the clients as malicious attackers. These attackers are presumed to conduct full-knowledge attacks, encompassing both backdoor and Trim attacks. In the backdoor attack, a 4×4 white pixel trigger is placed at the bottom right corner, with label 0 targeted as the specific class for all datasets. For the Trim attack, we disrupted and altered the parameters uploaded by malicious clients. Specifically, 10\% of the total model parameters were selected randomly to either receive Gaussian noise or be replaced with random numbers.


\textbf{Recovery settings.} In Crab, $\alpha$ is set to 10\% for generating time window, and the number of $\lambda$ is set to 60\% for round selection in buffer window. In client selection, the number of $\delta$ is set to 70\% of all clients participating in FL. It is important to note that during the FL process, we are unaware of which clients are malicious. There is a possibility of selecting malicious clients. Hence, in the subsequent recovery process, we exclude these malicious clients, selecting only the normal clients for the recovery process. For adaptive rollback, we set the sensitivity threshold $\beta$ to $0.3$.


\subsubsection{Compared Methods} We benchmark Crab against three recovery methods: (i) train-from-scratch, where after identifying malicious clients, the global model is retrained from scratch using only benign clients, (ii) FedEraser\cite{liu2021federaser}, which is similar to Crab but differs in that the server stores historical information at fixed interval rounds rather than based on importance, and (iii) FedRecover\cite{cao2023fedrecover}, which shifts the update renewal tasks from clients to the server, incorporating estimation and periodic correction.

\subsubsection{Evaluation Metrics} We employ various metrics to assess recovery performance, including (i) Test Accuracy (TA), which measures the proportion of test inputs correctly predicted by the global model, (ii) Backdoor Attack Success Rate (ASR), defined as the proportion of predictions classified as the target label in a dataset with an embedded backdoor trigger, (iii) Membership Inference Success Rate (MISR), introduced in \cite{shokri2017membership}, used to infer the data membership from malicious clients and thus evaluate recovery efficacy, and (iv) Running Time, which means the duration of each recovery round.

\subsubsection{Datasets} The efficiency and effectiveness of our framework is validated across datasets with a range of low and high resolution, including MNIST \cite{deng2012mnist}, Fashion-MNIST \cite{xiao2017fashion}, and CIFAR-10 \cite{krizhevsky2009learning}.

\subsection{Evaluation Results}

\begin{table}[!t]
    \centering
    \caption{Comparison of memory consumption and computation cost between baselines}
    \begin{tabular}{lll}
     \toprule
         & Memory consumption & Computation Cost\\
    \midrule
     Retrain    & $O(CMT)$ & $O(T)$\\
     FedEraser  & $O(CM\lfloor \frac{T}{\triangle t} \rfloor))$ & $O(\lfloor \frac{T}{\triangle t} \rfloor)$ \\
     FedRecover & $O(CMT)$  & $O(T_w+T_f+\lfloor \frac{(T-T_w-T_f)}{T_c} \rfloor$)  \\
     Crab  & \textbf{$O( \lambda \delta CMT)$} & \textbf{$O(\lambda T-j^\star)$}\\
     \bottomrule
    \end{tabular}  
    \label{tab:computation_memory}
\end{table}

\begin{figure}[t]
    \centering
    \includegraphics[width=1\linewidth]{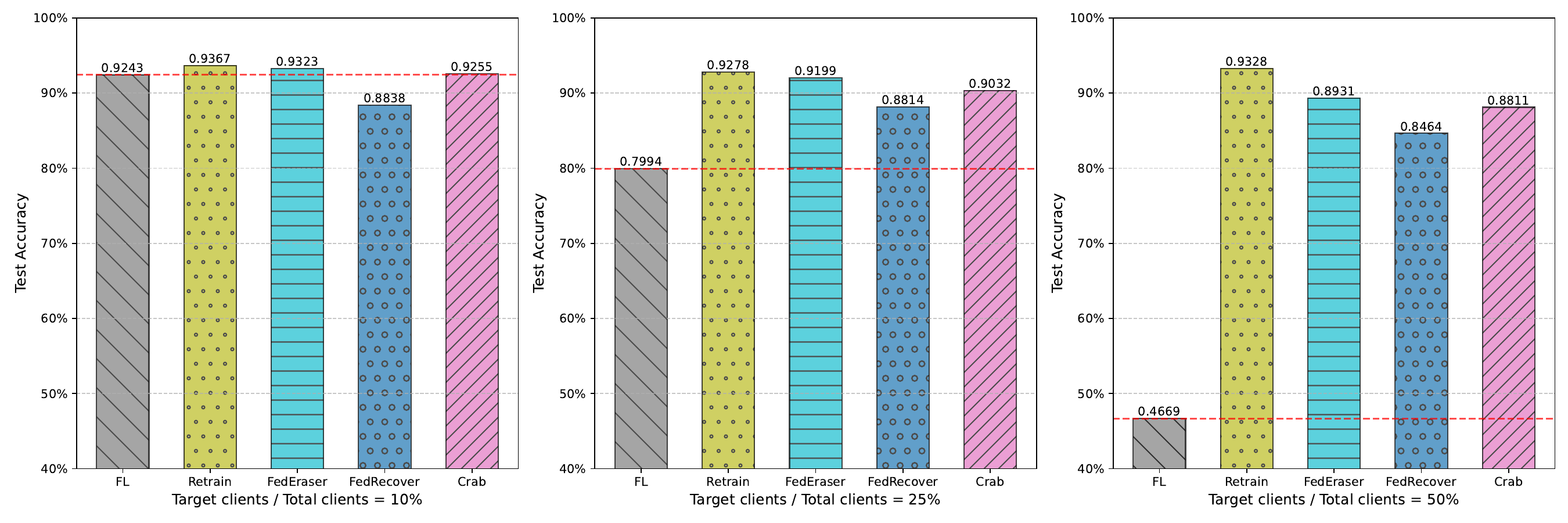}
    \caption{The test accuracy on MNIST dataset after recovering from backdoor attack when attack intensity is at 10\%, 25\% and 50\% respectively.}
    \label{fig:mnist_test_accuracy}
\end{figure}

\begin{table}[t]
\centering
\caption{Running time (sec) of each recovery round}
\label{tab:running time}
\resizebox{0.9\columnwidth}{!}{%
\begin{tabular}{@{}ccccc@{}}
\toprule
              & Retrain & FedEraser & FedRecover & Crab \\ \midrule
MNIST         & 1.47             & 1.56               & 2.67                & \textbf{0.68}          \\
Fashion-MNIST & 7.25             & 5.51               & 8.93                & \textbf{4.68}          \\
CIFAR-10      & 12.14            & 12.06              & 15.14               & \textbf{11.64}         \\ \bottomrule
\end{tabular}%
}
\end{table}
\textbf{Memory consumption and computation cost.}
To realize the model recovery, the server is required to store the information during the federated learning, which bring the memory consumption for the server. We assume that there are $M$ parameters in the local or global model. Train-from-scratch and FedRecover require server to store each round information from all client. Hence, the memory consumption for these two methods is the highest among the four methods, which can be $O(CMT)$, where $C$ is the total number of clients and $T$ is the total rounds. FedEraser stores each client's information every $\triangle t$ round so that the cost can be $O(CM\lfloor \frac{T}{\triangle t} \rfloor))$ to reduce the memory cost. In Crab, the cost depends on the selective stored rounds and clients. Therefore, the memory consumption for the server can be $O(\lambda \delta CMT)$. 

When to recover the poisoned model, the clients are required to compute the model update, which bring the computational cost to the client.
Train-from-scratch requires each client to re-compute model update in each round. Thus, the average computational cost per client is $O(T)$. FedEraser only asks each clients to calibrate the model for the selected round, which occur the computational cost $O(\lfloor \frac{T}{\triangle t} \rfloor)$. In FedRecover, the cost is $O(T_w+T_f+\lfloor \frac{(T-T_w-T_f)}{T_c} \rfloor)$, relying on the warm-up rounds $T_w$, the periodic correction $T_c$ and final tuning rounds $T_f$. In Crab, the cost depends on the selective stored round and roll-back round. Therefore, the computational cost for clients who participate the recovery process can be $O(\lambda T - j^\star)$. It is worth noting that not all client participate the recovery process.
The detail memory consumption for the server and computational cost for the clients is shown in Table \ref{tab:computation_memory}.

\textbf{Recovery speed.}
In our experiments, we evaluate the convergence rates of different recovery methods across several datasets, as shown in \cref{fig:loss_descent}. Our observations reveal that by selectively storing clients and utilizing historical global information, we can match the convergence speed of FedEraser, which uses small intervals for updates. Despite FedEraser theoretically setting an upper limit for our method due to its practical implementation with an interval of one, our method, Crab, closely aligns with FedEraser convergence curve after 20 rounds. Conversely, FedRecover, which employs the L-BFGS algorithm for calculating the Hessian matrix, not only increases the computation time per iteration but also shows significant instability, indicated by frequent extreme values in its convergence curve, necessitating periodic correction. Besides, we calculate the running time of each recovery round in various datasets, as shown in \cref{tab:running time}. Similarly, Crab demonstrates a clear advantage in terms of running time, which shows the superiority of selective storage strategy in improving the efficiency of recovery process.

Our adaptive rollback strategy also contributes to efficiency by eliminating the need to start from scratch each time. This approach proves particularly effective when dealing with varying proportions of malicious users. For instance, when the proportion is low, the rollback can commence from the 5th to 7th rounds, conserving computational resources. In cases where malicious users comprise up to 50\%, the rollback typically adjusts to start from the 2nd to 3rd rounds.

\begin{figure}[t]
    \centering
    \includegraphics[width=\linewidth]{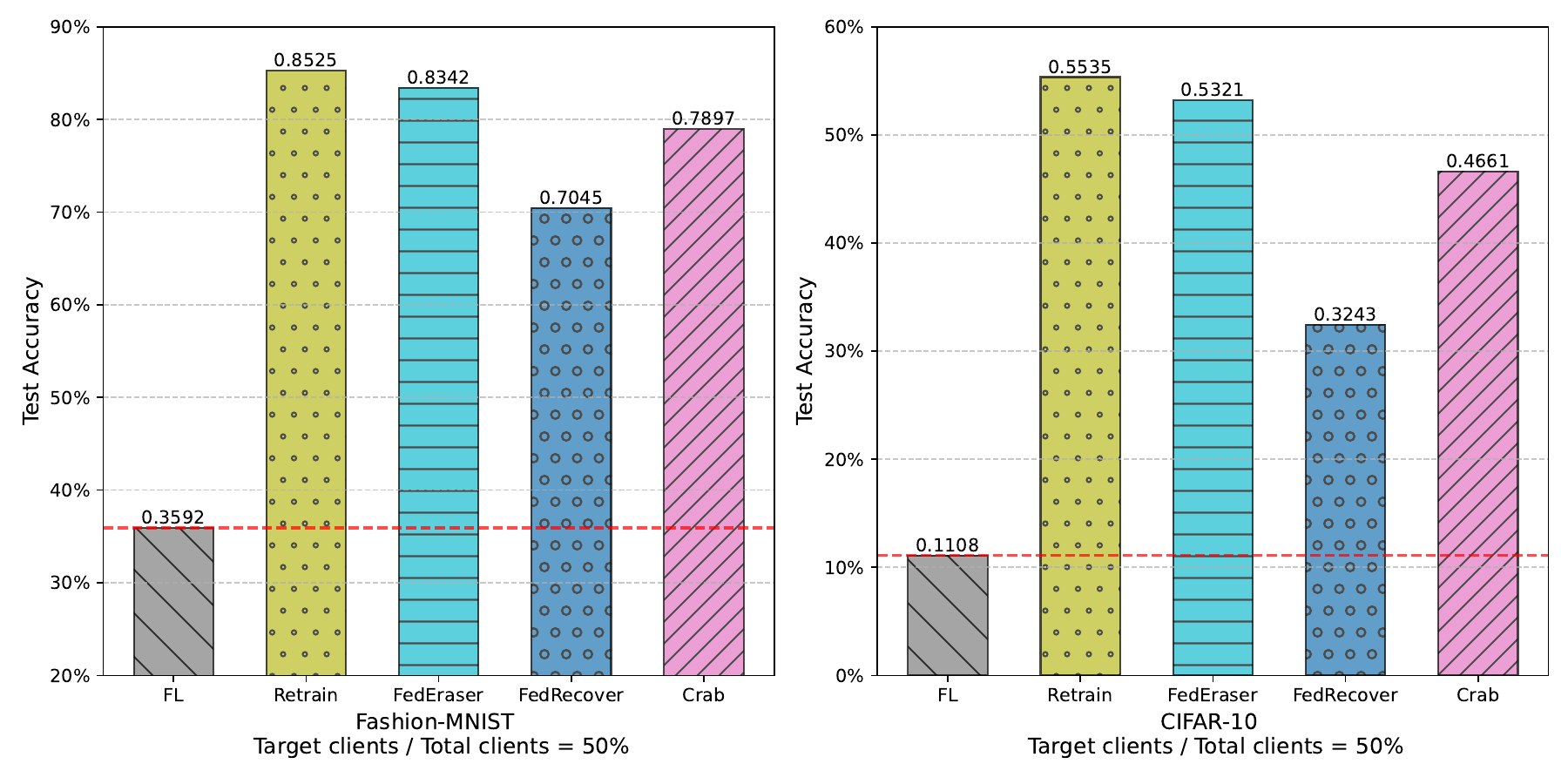}
    \caption{The test accuracy on Fashion-MNIST and CIFAR-10 after recovering from backdoor attack when attack intensity is at 50\%.}
    \label{fig:fmnist_cifar10_test_accuracy}
\end{figure}


\textbf{Accuracy on test data.}
We first present the recovery performance results under various backdoor attack ratios, shown in \cref{fig:mnist_test_accuracy}. We find that for backdoor attacks when the attack intensity is at 10\%, the impact on the FedAvg algorithm is limited. However, when the attack ratio increases to 50\%, a significant drop in test set accuracy is observed, falling to only 46.69\%. At this point, using our recovery algorithm Crab, the accuracy of the recovered global model can be restored to 88.11\%, which is only a 5\% decrease compared to the 93.28\% achieved with retraining. In the other two datasets, as shown in \cref{fig:fmnist_cifar10_test_accuracy}, we directly apply the strongest attack ratio and observe similar results: the Crab algorithm demonstrates superior recovery performance on both the Fashion-MNIST and CIFAR-10 datasets, with accuracy of 78.97\% and 46.61\% respectively, significantly surpassing the FedRecover algorithm.

\begin{figure}
    \centering
    \includegraphics[width=\linewidth]{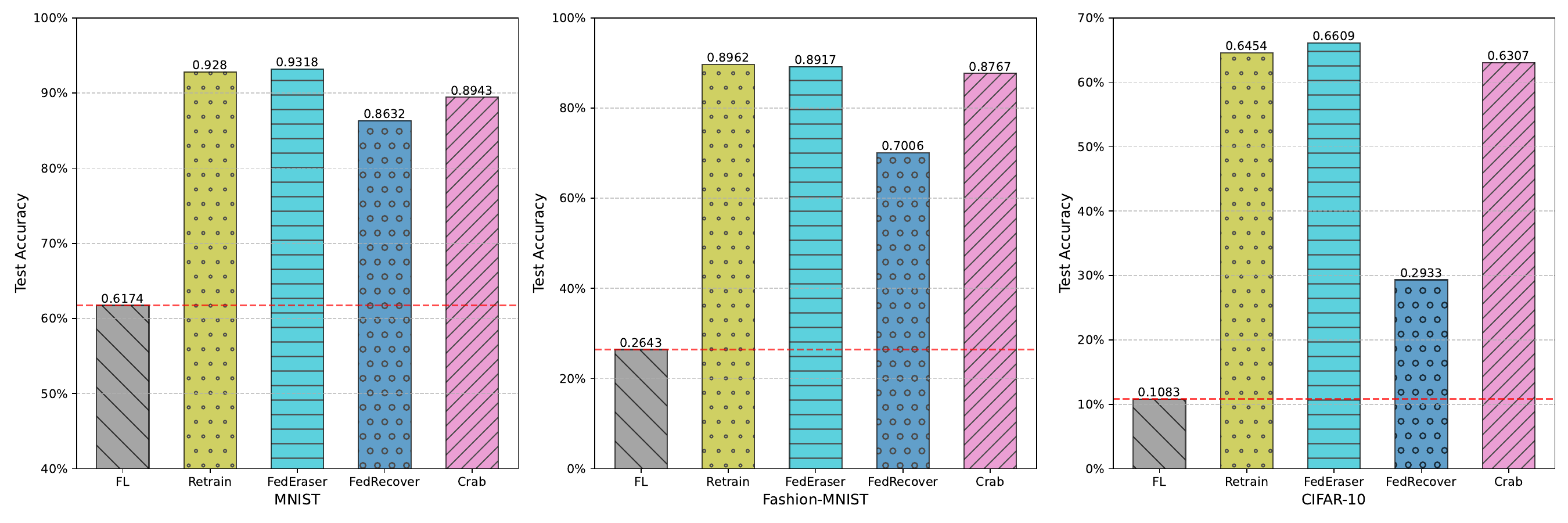}
    \caption{The test accuracy on MNIST, Fashion-MNIST and CIFAR-10 after recovering from trim attack.}
    \label{fig:trim_attack_test_accuracy}
\end{figure}

Regarding trim attacks, we arrive at a similar conclusion, shown in \cref{fig:trim_attack_test_accuracy}. The Crab algorithm achieved an accuracy of 89.43\%, 87.67\%, and 63.07\% on the three datasets, respectively. Compared to the performance of the poisoned global model, our method can effectively recover the model from attack.

\begin{figure}
    \centering
    \includegraphics[width=1\linewidth]{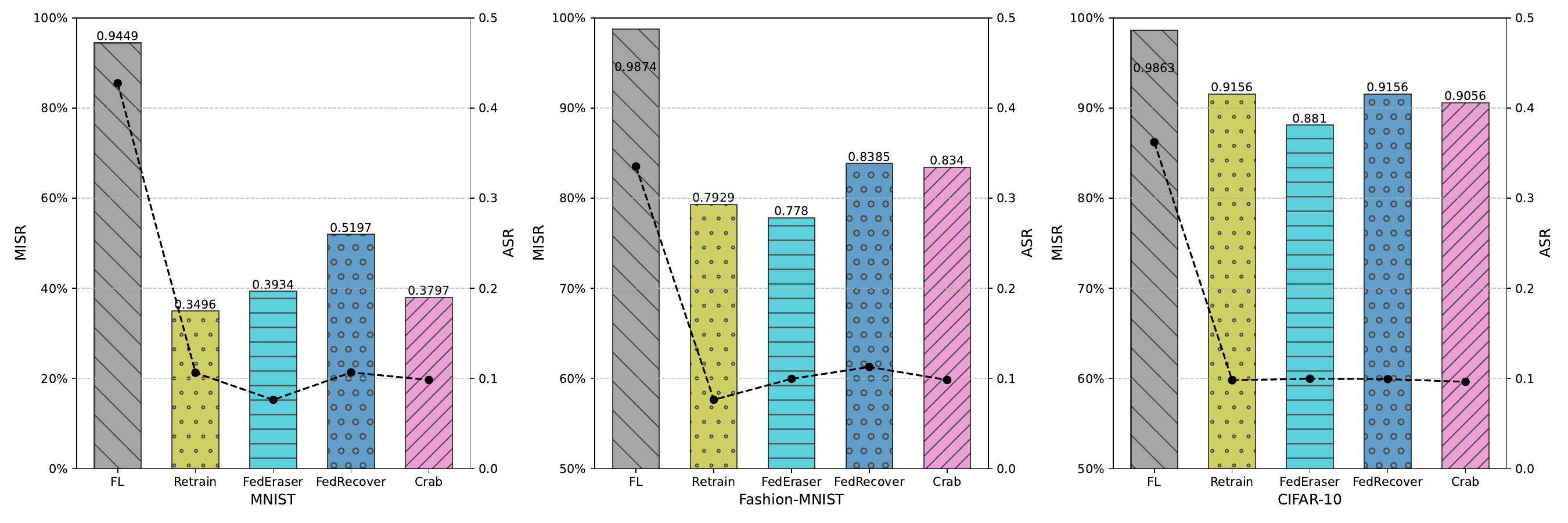}
    \caption{The MISR (presented in the bar chart) and ASR (presented in the line chart) on MNIST, Fashion-MNIST and CIFAR-10 after recovering from backdoor attack.}
    \label{fig:backdoor_asr}
\end{figure}

\textbf{Resilience to poisoning attacks.}
When evaluating a recovery algorithm, it is essential to consider not only its high accuracy on test datasets but also its ability to prevent malicious outputs. For instance, in the case of backdoor attacks, it is crucial to assess whether the global model still exhibits predetermined results when fed data containing a trigger. Our findings, illustrated in the \cref{fig:backdoor_asr}, show that the Attack Success Rate (ASR) of the recovered global models has decreased to around 10\%. It is important to note that the target client's predefined output (with the trigger) is set to 0 so roughly 10\% of the data is thus correctly labelled. 10\% ASR suggests that the attack is almost ineffective against the recovered global model.

Furthermore, we evaluate whether the datasets of the target clients are still part of the recovered model's training base by using membership inference attacks. \cref{fig:backdoor_asr} reveals a substantial decrease in the Membership Inference Success Rate (MISR) compared to Federated Learning (FL). For example, on the MNIST dataset, the MISR is down to 37.97\%. This indicates the effectiveness of our recovery framework.

\subsection{Analysis}

\begin{figure}[t]
    \centering
    \includegraphics[width=1\linewidth]{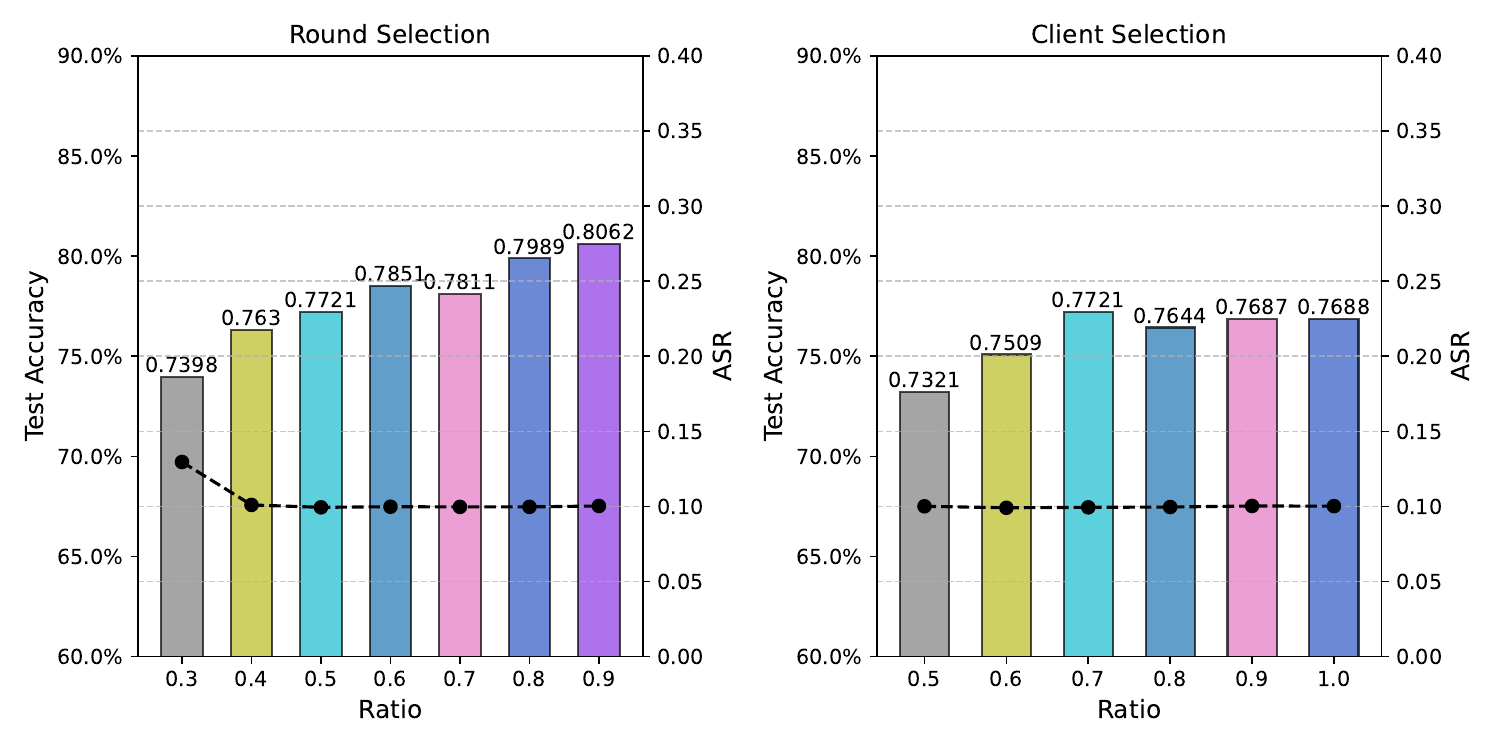}
    \caption{The test accuracy (presented in the bar chart)  and ASR (presented in the line chart) under different round and client selection ratio.}
    \label{fig:analysis_select}
\end{figure}

\subsubsection{The impact of selection rate} For the selective information storage strategy proposed for Crab, which includes two hyper parameters: the proportion $\lambda$ of round selection within a certain buffer window, and client selection rate $\delta$ selected in each chosen round. In the \cref{fig:analysis_select}, we conducted ablation experiments for these two parameters separately. 

Firstly, for the round selection's ratio $\lambda$, experiments show that using a lower ratio, such as 0.3, results in the lowest accuracy on the test set with the attack success rate close to 15\%. When gradually increase the proportion of storage rounds, it leads to consistent improvements in test accuracy. We balance the performance of the recovery model and the storage efficiency. We choose 0.6 as the proportion for round selection. Under this circumstance, the test accuracy can reach 78.51\%, and the attack success rate is also low.

After round selection, we choose specific clients with greater contributions in those rounds. We evaluate the recovery test accuracy at different select ratios of clients. When the ratio is increased to 0.8, there is a trend of decline of test accuracy. Therefore, we select a ratio of 0.7 as the proportion for client selection that not only maintains high testing performance but also meets the requirement of efficiency.

\begin{figure}[h]
    \centering
    \includegraphics[width=0.9\linewidth]{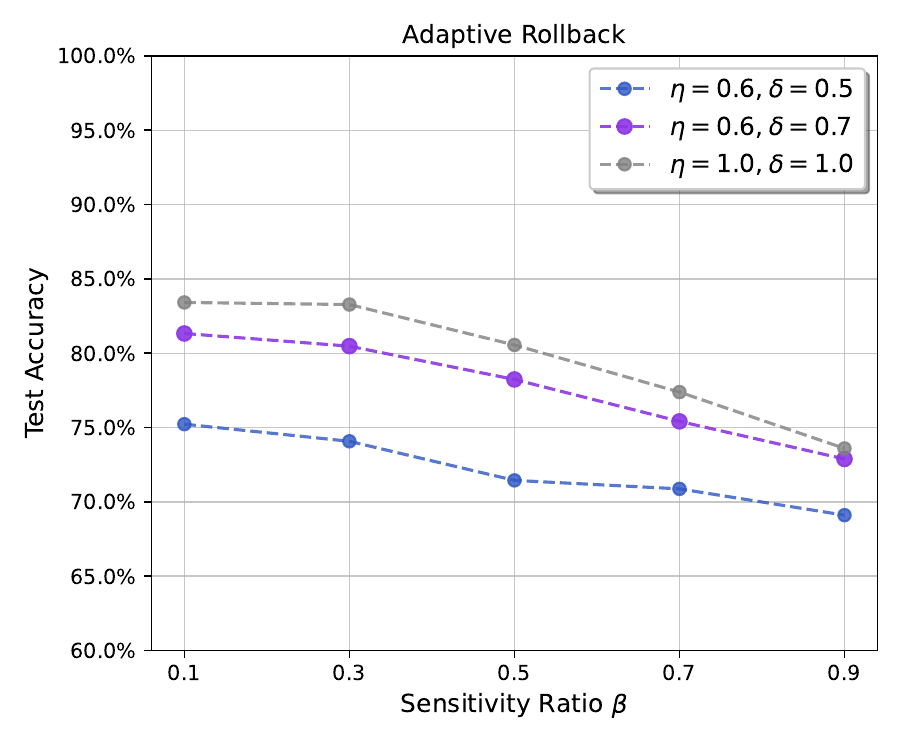}
    \caption{The test accuracy under different sensitivity bound.}
    \label{fig:Adaptive_Rollback}
\end{figure}

\subsubsection{The impact of sensitivity threshold} We analyze the performance of Crab under different sensitivity threshold ratio $\beta$, shown in \cref{fig:Adaptive_Rollback}. When the ratio is increased from 0.1 to 0.3, meaning the rollback starting point is moved further back, we do not observe a significant decrease in test accuracy, from 81.33\% to 80.48\%. This suggests that in the initial epochs, the global model is not significantly disturbed by the poisoning attack, and recovery does not need to start completely from the scratch. Therefore, selecting 0.3 as the sensitivity threshold ratio $\beta$ represents a reasonable trade-off between the capability and efficiency of model recovery. However, when the threshold is further increased to 0.9, the test accuracy shows a significant fluctuation and a noticeable decline, which indicates that the global model has already been injected with the attack triggers. If the model at this epoch is used as the roll back epoch, the performance of the recovered model will significantly decrease. Additionally, we further evaluate various results under different selection settings as shown in \cref{fig:Adaptive_Rollback}. Similarly, using the aforementioned parameters, namely $\lambda=0.6$ and $\delta=0.7$, is a relatively ideal balance point, which maintains both the effectiveness and efficiency of the recovery process.

\section{Conclusion} \label{sec:conclusions}
In summary, we introduce Crab, an efficient and certified method for recovering the global model from poisoning attacks, utilizing selective storage and adaptive rollback strategies. Theoretical analysis validates that the model difference between the global model recovered by Crab and the one recovered by training from scratch can be bounded under specific assumptions.
Our experiments affirm that Crab, while saving storage and computational costs, maintains the effectiveness and stability of model recovery. Furthermore, we explore the impact of Crab parameters on the trade-off between resource consumption and recovery accuracy. Crab surpasses other existing methods in both recovery speed and resource utilization.

\bibliographystyle{IEEEtran}
\bibliography{mybibliography}

\begin{thebibliography}{10}
\providecommand{\url}[1]{#1}
\csname url@samestyle\endcsname
\providecommand{\newblock}{\relax}
\providecommand{\bibinfo}[2]{#2}
\providecommand{\BIBentrySTDinterwordspacing}{\spaceskip=0pt\relax}
\providecommand{\BIBentryALTinterwordstretchfactor}{4}
\providecommand{\BIBentryALTinterwordspacing}{\spaceskip=\fontdimen2\font plus
\BIBentryALTinterwordstretchfactor\fontdimen3\font minus \fontdimen4\font\relax}
\providecommand{\BIBforeignlanguage}[2]{{%
\expandafter\ifx\csname l@#1\endcsname\relax
\typeout{** WARNING: IEEEtran.bst: No hyphenation pattern has been}%
\typeout{** loaded for the language `#1'. Using the pattern for}%
\typeout{** the default language instead.}%
\else
\language=\csname l@#1\endcsname
\fi
#2}}
\providecommand{\BIBdecl}{\relax}
\BIBdecl

\bibitem{kairouz2021advances}
P.~Kairouz, H.~B. McMahan, B.~Avent, A.~Bellet, M.~Bennis, A.~N. Bhagoji, K.~Bonawitz, Z.~Charles, G.~Cormode, R.~Cummings \emph{et~al.}, ``Advances and open problems in federated learning,'' \emph{Foundations and Trends{\textregistered} in Machine Learning}, 2021.

\bibitem{zhang2021survey}
C.~Zhang, Y.~Xie, H.~Bai, B.~Yu, W.~Li, and Y.~Gao, ``A survey on federated learning,'' \emph{Knowledge-Based Systems}, vol. 216, p. 106775, 2021.

\bibitem{liu2022privacy}
Z.~Liu, J.~Guo, W.~Yang, J.~Fan, K.-Y. Lam, and J.~Zhao, ``Privacy-preserving aggregation in federated learning: A survey,'' \emph{IEEE Transactions on Big Data}, 2022.

\bibitem{li2020federated}
T.~Li, A.~K. Sahu, A.~Talwalkar, and V.~Smith, ``Federated learning: Challenges, methods, and future directions,'' \emph{IEEE signal processing magazine}, vol.~37, no.~3, pp. 50--60, 2020.

\bibitem{hang2023privacy}
C.-N. Hang, Y.-Z. Tsai, P.-D. Yu, J.~Chen, and C.-W. Tan, ``Privacy-enhancing digital contact tracing with machine learning for pandemic response: A comprehensive review,'' \emph{Big Data and Cognitive Computing}, vol.~7, no.~2, p. 108, 2023.

\bibitem{shejwalkar2022back}
V.~Shejwalkar, A.~Houmansadr, P.~Kairouz, and D.~Ramage, ``Back to the drawing board: A critical evaluation of poisoning attacks on production federated learning,'' in \emph{2022 IEEE Symposium on Security and Privacy (SP)}.\hskip 1em plus 0.5em minus 0.4em\relax IEEE, 2022, pp. 1354--1371.

\bibitem{jebreel2023fl}
N.~M. Jebreel and J.~Domingo-Ferrer, ``Fl-defender: Combating targeted attacks in federated learning,'' \emph{Knowledge-Based Systems}, vol. 260, p. 110178, 2023.

\bibitem{bagdasaryan2020backdoor}
E.~Bagdasaryan, A.~Veit, Y.~Hua, D.~Estrin, and V.~Shmatikov, ``How to backdoor federated learning,'' in \emph{International conference on artificial intelligence and statistics}.\hskip 1em plus 0.5em minus 0.4em\relax PMLR, 2020, pp. 2938--2948.

\bibitem{zhang2023fltracer}
X.~Zhang, Q.~Liu, Z.~Ba, Y.~Hong, T.~Zheng, F.~Lin, L.~Lu, and K.~Ren, ``Fltracer: Accurate poisoning attack provenance in federated learning,'' \emph{arXiv preprint arXiv:2310.13424}, 2023.

\bibitem{blanchard2017machine}
P.~Blanchard, E.~M. El~Mhamdi, R.~Guerraoui, and J.~Stainer, ``Machine learning with adversaries: Byzantine tolerant gradient descent,'' \emph{Advances in neural information processing systems}, vol.~30, 2017.

\bibitem{lyu2020threats}
L.~Lyu, H.~Yu, and Q.~Yang, ``Threats to federated learning: A survey,'' \emph{arXiv preprint arXiv:2003.02133}, 2020.

\bibitem{zhang2022fldetector}
Z.~Zhang, X.~Cao, J.~Jia, and N.~Z. Gong, ``Fldetector: Defending federated learning against model poisoning attacks via detecting malicious clients,'' in \emph{Proceedings of the 28th ACM SIGKDD Conference on Knowledge Discovery and Data Mining}, 2022, pp. 2545--2555.

\bibitem{li2020learning}
S.~Li, Y.~Cheng, W.~Wang, Y.~Liu, and T.~Chen, ``Learning to detect malicious clients for robust federated learning,'' \emph{arXiv preprint arXiv:2002.00211}, 2020.

\bibitem{shen2016auror}
S.~Shen, S.~Tople, and P.~Saxena, ``Auror: Defending against poisoning attacks in collaborative deep learning systems,'' in \emph{Proceedings of the 32nd Annual Conference on Computer Security Applications}, 2016, pp. 508--519.

\bibitem{cao2023fedrecover}
X.~Cao, J.~Jia, Z.~Zhang, and N.~Z. Gong, ``Fedrecover: Recovering from poisoning attacks in federated learning using historical information,'' in \emph{2023 IEEE Symposium on Security and Privacy (SP)}.\hskip 1em plus 0.5em minus 0.4em\relax IEEE, 2023, pp. 1366--1383.

\bibitem{liu2021federaser}
G.~Liu, X.~Ma, Y.~Yang, C.~Wang, and J.~Liu, ``Federaser: Enabling efficient client-level data removal from federated learning models,'' in \emph{2021 IEEE/ACM 29th International Symposium on Quality of Service (IWQOS)}.\hskip 1em plus 0.5em minus 0.4em\relax IEEE, 2021, pp. 1--10.

\bibitem{liu2023survey}
Z.~Liu, Y.~Jiang, J.~Shen, M.~Peng, K.-Y. Lam, and X.~Yuan, ``A survey on federated unlearning: Challenges, methods, and future directions,'' \emph{arXiv preprint arXiv:2310.20448}, 2023.

\bibitem{halimi2022federated}
A.~Halimi, S.~R. Kadhe, A.~Rawat, and N.~B. Angel, ``Federated unlearning: How to efficiently erase a client in fl?'' in \emph{International Conference on Machine Learning}, 2022.

\bibitem{liu2022right}
Y.~Liu, L.~Xu, X.~Yuan, C.~Wang, and B.~Li, ``The right to be forgotten in federated learning: An efficient realization with rapid retraining,'' in \emph{IEEE INFOCOM 2022-IEEE Conference on Computer Communications}.\hskip 1em plus 0.5em minus 0.4em\relax IEEE, 2022, pp. 1749--1758.

\bibitem{mcmahan2017communication}
B.~McMahan, E.~Moore, D.~Ramage, S.~Hampson, and B.~A. y~Arcas, ``Communication-efficient learning of deep networks from decentralized data,'' in \emph{Artificial intelligence and statistics}.\hskip 1em plus 0.5em minus 0.4em\relax PMLR, 2017, pp. 1273--1282.

\bibitem{li2020federated2}
T.~Li, A.~K. Sahu, M.~Zaheer, M.~Sanjabi, A.~Talwalkar, and V.~Smith, ``Federated optimization in heterogeneous networks,'' \emph{Proceedings of Machine learning and systems}, vol.~2, pp. 429--450, 2020.

\bibitem{zhang2023symmetric}
Y.~Zhang, C.~Li, and C.~W. Tan, ``On symmetric multilevel diversity coding system with linear computations,'' \emph{IEEE Communications Letters}, 2023.

\bibitem{wang2020tackling}
J.~Wang, Q.~Liu, H.~Liang, G.~Joshi, and H.~V. Poor, ``Tackling the objective inconsistency problem in heterogeneous federated optimization,'' \emph{Advances in neural information processing systems}, vol.~33, pp. 7611--7623, 2020.

\bibitem{karimireddy2020scaffold}
S.~P. Karimireddy, S.~Kale, M.~Mohri, S.~Reddi, S.~Stich, and A.~T. Suresh, ``Scaffold: Stochastic controlled averaging for federated learning,'' in \emph{International conference on machine learning}.\hskip 1em plus 0.5em minus 0.4em\relax PMLR, 2020, pp. 5132--5143.

\bibitem{fang2020local}
M.~Fang, X.~Cao, J.~Jia, and N.~Gong, ``Local model poisoning attacks to $\{$Byzantine-Robust$\}$ federated learning,'' in \emph{29th USENIX security symposium (USENIX Security 20)}, 2020, pp. 1605--1622.

\bibitem{li2022backdoor}
Y.~Li, Y.~Jiang, Z.~Li, and S.-T. Xia, ``Backdoor learning: A survey,'' \emph{IEEE Transactions on Neural Networks and Learning Systems}, 2022.

\bibitem{gu2017badnets}
T.~Gu, B.~Dolan-Gavitt, and S.~Garg, ``Badnets: Identifying vulnerabilities in the machine learning model supply chain,'' \emph{arXiv preprint arXiv:1708.06733}, 2017.

\bibitem{liu2018trojaning}
Y.~Liu, S.~Ma, Y.~Aafer, W.-C. Lee, J.~Zhai, W.~Wang, and X.~Zhang, ``Trojaning attack on neural networks,'' in \emph{25th Annual Network And Distributed System Security Symposium (NDSS 2018)}.\hskip 1em plus 0.5em minus 0.4em\relax Internet Soc, 2018.

\bibitem{li2020invisible}
S.~Li, M.~Xue, B.~Z.~H. Zhao, H.~Zhu, and X.~Zhang, ``Invisible backdoor attacks on deep neural networks via steganography and regularization,'' \emph{IEEE Transactions on Dependable and Secure Computing}, vol.~18, no.~5, pp. 2088--2105, 2020.

\bibitem{bagdasaryan2021blind}
E.~Bagdasaryan and V.~Shmatikov, ``Blind backdoors in deep learning models,'' in \emph{30th USENIX Security Symposium (USENIX Security 21)}, 2021, pp. 1505--1521.

\bibitem{saha2020hidden}
A.~Saha, A.~Subramanya, and H.~Pirsiavash, ``Hidden trigger backdoor attacks,'' in \emph{Proceedings of the AAAI conference on artificial intelligence}, vol.~34, no.~07, 2020, pp. 11\,957--11\,965.

\bibitem{bhagoji2019analyzing}
A.~N. Bhagoji, S.~Chakraborty, P.~Mittal, and S.~Calo, ``Analyzing federated learning through an adversarial lens,'' in \emph{International Conference on Machine Learning}.\hskip 1em plus 0.5em minus 0.4em\relax PMLR, 2019, pp. 634--643.

\bibitem{xie2019dba}
C.~Xie, K.~Huang, P.-Y. Chen, and B.~Li, ``Dba: Distributed backdoor attacks against federated learning,'' in \emph{International conference on learning representations}, 2019.

\bibitem{shejwalkar2021manipulating}
V.~Shejwalkar and A.~Houmansadr, ``Manipulating the byzantine: Optimizing model poisoning attacks and defenses for federated learning,'' in \emph{NDSS}, 2021.

\bibitem{baruch2019little}
G.~Baruch, M.~Baruch, and Y.~Goldberg, ``A little is enough: Circumventing defenses for distributed learning,'' \emph{Advances in Neural Information Processing Systems}, vol.~32, 2019.

\bibitem{cover1999elements}
T.~M. Cover, \emph{Elements of information theory}.\hskip 1em plus 0.5em minus 0.4em\relax John Wiley \& Sons, 1999.

\bibitem{fang2022robust}
X.~Fang and M.~Ye, ``Robust federated learning with noisy and heterogeneous clients,'' in \emph{Proceedings of the IEEE/CVF Conference on Computer Vision and Pattern Recognition}, 2022, pp. 10\,072--10\,081.

\bibitem{zhang2022personalized}
X.~Zhang, Y.~Li, W.~Li, K.~Guo, and Y.~Shao, ``Personalized federated learning via variational bayesian inference,'' in \emph{International Conference on Machine Learning}.\hskip 1em plus 0.5em minus 0.4em\relax PMLR, 2022, pp. 26\,293--26\,310.

\bibitem{cao2021fltrust}
X.~Cao, M.~Fang, J.~Liu, and N.~Gong, ``Fltrust: Byzantine-robust federated learning via trust bootstrapping,'' in \emph{Proceedings of NDSS}, 2021.

\bibitem{liu2021privacy}
X.~Liu, H.~Li, G.~Xu, Z.~Chen, X.~Huang, and R.~Lu, ``Privacy-enhanced federated learning against poisoning adversaries,'' \emph{IEEE Transactions on Information Forensics and Security}, vol.~16, pp. 4574--4588, 2021.

\bibitem{ma2022shieldfl}
Z.~Ma, J.~Ma, Y.~Miao, Y.~Li, and R.~H. Deng, ``Shieldfl: Mitigating model poisoning attacks in privacy-preserving federated learning,'' \emph{IEEE Transactions on Information Forensics and Security}, vol.~17, pp. 1639--1654, 2022.

\bibitem{dwork2014algorithmic}
C.~Dwork, A.~Roth \emph{et~al.}, ``The algorithmic foundations of differential privacy,'' \emph{Foundations and Trends{\textregistered} in Theoretical Computer Science}, vol.~9, no. 3--4, pp. 211--407, 2014.

\bibitem{fraboni2022sequential}
Y.~Fraboni, R.~Vidal, L.~Kameni, and M.~Lorenzi, ``Sequential informed federated unlearning: Efficient and provable client unlearning in federated optimization,'' \emph{arXiv preprint arXiv:2211.11656}, 2022.

\bibitem{guo2019certified}
C.~Guo, T.~Goldstein, A.~Hannun, and L.~Van Der~Maaten, ``Certified data removal from machine learning models,'' \emph{arXiv preprint arXiv:1911.03030}, 2019.

\bibitem{zhou2021communication}
Y.~Zhou, Q.~Ye, and J.~Lv, ``Communication-efficient federated learning with compensated overlap-fedavg,'' \emph{IEEE Transactions on Parallel and Distributed Systems}, vol.~33, no.~1, pp. 192--205, 2021.

\bibitem{beck2017first}
A.~Beck, \emph{First-order methods in optimization}.\hskip 1em plus 0.5em minus 0.4em\relax SIAM, 2017.

\bibitem{shokri2017membership}
R.~Shokri, M.~Stronati, C.~Song, and V.~Shmatikov, ``Membership inference attacks against machine learning models,'' in \emph{2017 IEEE symposium on security and privacy (SP)}.\hskip 1em plus 0.5em minus 0.4em\relax IEEE, 2017, pp. 3--18.

\bibitem{deng2012mnist}
L.~Deng, ``The mnist database of handwritten digit images for machine learning research [best of the web],'' \emph{IEEE signal processing magazine}, vol.~29, no.~6, pp. 141--142, 2012.

\bibitem{xiao2017fashion}
H.~Xiao, K.~Rasul, and R.~Vollgraf, ``Fashion-mnist: a novel image dataset for benchmarking machine learning algorithms,'' \emph{arXiv preprint arXiv:1708.07747}, 2017.

\bibitem{krizhevsky2009learning}
A.~Krizhevsky, G.~Hinton \emph{et~al.}, ``Learning multiple layers of features from tiny images,'' 2009.

\end{thebibliography}

\clearpage

\appendices

\section{Proof of Theorem \ref{theo_1}}
\label{Proof of Theorem}
In this part, we show that the difference between the global model recovered by Crab and that recovered by train-from-scratch can be bounded, i.e., $\Vert \tilde{M}_r - M_{\tau}\Vert$ is bounded.

Recall that the global model recovered by Crab is updated as follows:
\begin{equation}
   \tilde{M}_{r+1} = \tilde{M}_r- \eta \tilde{\boldsymbol{H}}_r,
\end{equation}
where $ \tilde{\boldsymbol{H}}_r = \sum_{c=1}^{X_{j^\star+r}} \sigma_r^c  \cdot \hat{\boldsymbol g}_r^c  $, and $\sigma_r^c = \frac{ |D_c| }{|D_{t_{j^\star + r }}^{-}|}  \cdot \frac{\Vert \boldsymbol g_{t_{j^\star + r }}^c \Vert}{\Vert \hat{\boldsymbol g}_{r}^c \Vert} $.

Let $\boldsymbol{h}_\tau^c$ denote the model update from client $c$ in round $\tau$ of train-from-scratch.
The global model recovered by train-from-scratch is updated as follows:
\begin{equation}
    M_{\tau+1}=M_{\tau}-\eta \boldsymbol{H}_{\tau},
\end{equation}
where $\boldsymbol{H}_{\tau}= \sum_i^{n-m} \frac{|D_i|}{|D^{-}|} \cdot \boldsymbol{h}_\tau^c$. 

Considering the round selection by Crab, we can bound the difference between the global model recovered by Crab in round $t$ and that recovered by train-from-scratch in round $\tau$: 
\begin{equation}
    \begin{aligned}
        &\Vert \tilde{M}_r -  M_{\tau} \Vert \\
        = &\Vert (\tilde{M}_r - \tilde M_0) -  (M_{\tau} - \tilde M_0)\Vert \\
        \leq & \Vert \tilde{M}_r - \tilde M_0\Vert+ \Vert M_{\tau} - \tilde M_0\Vert,
    \end{aligned}
\end{equation}
where $\tau = \lceil r \cdot \frac{T}{(T^\prime-j^\star)} \rceil $, and $\lceil \cdot \rceil$ is the ceiling function. In the recovery process, the initial model is the optimal roll-back model, which is used in both Crab and train-from-scratch method. Thus, $\tilde M_0 = \bar M_{t_{j^\star}}$. Then, according to the triangle inequality and the homogeneity, we can compute:
\begin{equation}
    \begin{aligned}
        &\Vert \tilde{M}_r - \tilde M_0\Vert 
        = \Vert  -\eta \sum_{r=0}^{r-1} \tilde{\boldsymbol{H}}_r \Vert 
        \leq  \eta \sum_{r=0}^{r-1} \Vert \tilde{\boldsymbol{H}}_r  \Vert,
    \end{aligned}
\end{equation}
and
\begin{equation}
    \begin{aligned}
        &\Vert M_{\tau} - \tilde M_0\Vert 
        = \Vert  -\eta \sum_{{\tau}=0}^{{\tau}-1} \boldsymbol{H}_{\tau}  \Vert 
        \leq  \eta \sum_{{\tau}=0}^{{\tau}-1} \Vert \boldsymbol{H}_{\tau} \Vert.
    \end{aligned}
\end{equation}

Firstly, we compute the $\Vert M_{\tau} - \tilde M_0\Vert $.
Recall that in the Assumption \ref{ass_smooth}, by the smoothness of function $F$ and update rule, we have:
\begin{equation}
    \begin{aligned}
        &F(M_{\tau+1}) \\
        \leq &F(M_{\tau}) + \nabla F(M_{\tau})^\top (M_{\tau+1} - M_{\tau}) + \frac{L}{2} \Vert M_{\tau+1} - M_{\tau} \Vert^2 \\
        = & F(M_{\tau}) - \eta \nabla F(M_{\tau})^\top \boldsymbol{H}_{\tau}  + \frac{L}{2} \eta^2 \Vert \boldsymbol{H}_{\tau} \Vert^2 .
    \end{aligned}
\end{equation}
According to the Assumption \ref{ass_gradient}, $\boldsymbol{H}_{\tau}$ is an unbiased estimation of $\nabla F(M_{\tau})$, and $\Vert f_c(M) \Vert \leq G$, we can have:
\begin{equation}
        \begin{aligned}
        F(M_{\tau+1}) 
        \leq & F(M_{\tau}) - \eta \Vert \nabla F(M_{\tau}) \Vert^2 + \frac{L}{2} \eta^2 G^2.
    \end{aligned}
\end{equation}
Hence, by rearranging the terms, we have:
\begin{equation}
    \eta \Vert \nabla F(M_{\tau}) \Vert^2 \leq  F(M_{\tau}) - F(M_{\tau+1}) + \frac{L}{2} \eta^2 G^2.
\end{equation}
When to iteration $\tau$ rounds, we have
\begin{equation}
\begin{aligned}
        \sum_{\tau=0}^{\tau-1} \eta \Vert \nabla F(M_{\tau}) \Vert^2 &\leq  F(\tilde M_0) - F(M_{\tau-1}) +  \frac{L}{2} \tau \eta^2 G^2 \\
        &\leq F(\tilde M_0) - F(M^*) +  \frac{L}{2} \tau \eta^2 G^2,
\end{aligned}
\end{equation}
where $M^*$ is the optimal solution for the function $F(M)$.
Hence, by applying the Cauchy–Schwarz inequality, we can compute
\begin{equation}
\begin{aligned}
       \Vert M_{\tau} - \tilde M_0\Vert
       &\leq  \eta \sum_{{\tau}=0}^{{\tau}-1} \Vert \boldsymbol{H}_{\tau}  \Vert \\
       \leq & \sqrt{ \sum_{\tau=0}^{\tau-1} \eta \sum_{\tau=0}^{\tau-1} \eta \Vert \nabla F(M_{\tau}) \Vert^2 } \\
       \leq &\sqrt{ \sum_{\tau=0}^{\tau-1} \eta [ F(\tilde M_0) - F(M^*) +  \frac{L}{2} \tau \eta^2 G^2 ]}.
\end{aligned}
\end{equation}

Then, we compute $\Vert \tilde{M}_r - \tilde M_{0} \Vert$.
Similarly, we can have 
\begin{equation}
    \eta \Vert \nabla F(\tilde M_{r}) \Vert^2 \leq  F(\tilde M_{r}) - F(\tilde M_{r+1}) + \frac{L}{2} \eta^2 G^2 (\sum_{c=1}^{X_{j^\star+r}} \sigma_r^c)^2.
\end{equation}
When to iteration $r$ rounds, we have
\begin{equation}
\begin{aligned}
        &\sum_{r=0}^{r-1} \eta \Vert \nabla F(\tilde M_{r}) \Vert^2 \\
        \leq & F(\tilde M_{0}) - F(M^*) + \frac{L}{2} \eta^2 G^2 \sum_{r=0}^{r-1} (\sum_{c=1}^{X_{j^\star+r}} \sigma_r^c)^2.
\end{aligned}
\end{equation}
For simplicity, we denote $\check{\sigma}_r = \sum_{c=1}^{X_{j^\star+r}} \sigma_r^c$. 
Then, we can get
\begin{equation}
\begin{aligned}
       &\Vert \tilde M_{r} - \tilde M_0\Vert 
        \leq  \eta \sum_{r=0}^{r-1} \Vert \tilde{\boldsymbol{H}}_r  \Vert \\
       \leq &\sqrt{ \sum_{r=0}^{r-1} \eta [ F(\tilde M_0) - F(M^*) +  \frac{L}{2} \eta^2 G^2 \sum_{r=0}^{r-1} \check{\sigma}_r^2]}.
\end{aligned}
\end{equation}
Finally, the difference between $\tilde{M}_r$ and $M_\tau$ can be:
\begin{equation}
\begin{aligned}
    &\Vert \tilde{M}_r -  M_{\tau} \Vert\\
    \leq & \Vert \tilde{M}_r - \tilde M_0\Vert+ \Vert M_{\tau} - \tilde M_0\Vert \\
    \leq & \sqrt{ \eta [ (F(\tilde M_0) - F(M^*))  ( r + \tau) + \frac{L}{2} \eta^2 G^2 (r\sum_{r=0}^{r-1} \check{\sigma}_r^2 +\tau) ]},
\end{aligned}
\end{equation}
where $\tau = \lceil r \cdot \frac{T}{(T^\prime-j^\star)} \rceil $, $\tilde M_0 = \bar M_{t_{j^\star}}$ is the initial model used in both Crab and train-from-scratch method, $M^*$ is the optimal solution for the function $F(M)$, $\check{\sigma}_r = \sum_{c=1}^{X_{j^\star+r}} \sigma_r^c$, and $\sigma_r^c = \frac{ |D_c| }{|D_{t_{j^\star + r }}^{-}|}  \cdot \frac{\Vert \boldsymbol g_{t_{j^\star + r }}^c \Vert}{\Vert \hat{\boldsymbol g}_{r}^c \Vert} $.


\end{document}